\documentclass[onecolumn, draftcls, 12 pt,  a4paper]{IEEEtran}  
\usepackage{graphicx}
\usepackage{color}
\usepackage{epsfig}
\usepackage{amsmath}
\usepackage{amssymb, amsfonts}
\usepackage{latexsym}
\usepackage{booktabs}

\newcommand{\boldgamma}{\mbox{\boldmath{$\gamma$}}}
\newcommand{\mean}[1]{\langle #1 \rangle}
\newcommand{\expectation}[2]{\mathbb{E}_{#2}\left[#1\right]}
\newcommand{\IX}{I_{\mathcal X}}
\newcommand{\popt}{p^{\rm opt}}
\newcommand{\pref}{p^{\rm ref}}
\newcommand{\lp}{\wp}
\newcommand{\boldlp}{\mbox{\boldmath{$\lp$}}}
\newcommand{\pst}{\mathbf{p}_{\rm st}}
\newcommand{\plt}{\mathbf{p}_{\rm lt}}
\newcommand{\peq}{\mathbf{p}_{\rm eq}}
\newcommand{\pltopt}{\mathbf{\popt_{\rm lt}}}

\newcommand{\ptw}{p^{\rm tw}}

\newcommand{\MMSE}{{\rm MMSE}_{\mathcal X}}
\newcommand{\Pout}{P_{\rm out}}
\newcommand{\Poutlt}{P_{\rm out}^{\rm lt}}
\newcommand{\ptwfill}{p^{\rm tw}}
\newcommand{\openone}{\leavevmode\hbox{\small1\normalsize\kern-.33em1}}

\newtheorem{theorem}{Theorem}
\newtheorem{lemma}{Lemma}
\newtheorem{proposition}{Proposition}

\newtheorem{corollary}{Corollary}

\title{Power Allocation for Discrete-Input Delay-Limited Fading Channels}

\author{Khoa D. Nguyen, Albert Guill\'en i F\`abregas and Lars
  K. Rasmussen\thanks{K. D. Nguyen and L. K. Rasmussen are with the Institute
    for Telecommunications Research, University of South Australia, Mawson Lakes Boulevard, Mawson Lakes 5095, South Australia, Australia,
    \tt e-mail: dangkhoa.nguyen@postgrads.unisa.edu.au,
    lars.rasmussen@unisa.edu.au.}\thanks{A. Guill\'en i F\`abregas is with the
    Department of Engineering, University of Cambridge, Trumpington Street, Cambridge CB2 1PZ, UK, \tt e-mail:
    guillen@ieee.org.}
    \thanks{Parts of this work will be presented in the 2007 IEEE Information Theory Workshop, Lake Tahoe, CA, USA, September 2007.}
    \thanks{This work has been supported by the Australian Research Council under ARC grants RN0459498 and DP0558861.}}

\begin{document}

\maketitle

\begin{abstract}
We consider power allocation algorithms for fixed-rate
transmission over Nakagami-$m$ non-ergodic block-fading channels
with perfect transmitter and receiver channel state information
and discrete input signal constellations, under both short- and
long-term power constraints. Optimal power allocation schemes are
shown to be direct applications of previous results in the
literature. We show that the SNR exponent of the optimal
short-term scheme is given by $m$ times the Singleton bound. We also
illustrate the significant gains available by employing long-term
power constraints. In particular, we analyze the optimal long-term
solution, showing that zero outage can be achieved provided that
the corresponding short-term SNR exponent with the same system parameters is strictly greater than one. Conversely, if
the short-term SNR exponent  is smaller than one, we show that zero outage
cannot be achieved. In this case, we derive the corresponding long-term SNR
exponent as a function of the Singleton bound. Due to the nature
of the expressions involved, the complexity of optimal schemes may
be prohibitive for system implementation. We therefore propose
simple sub-optimal power allocation schemes whose outage
probability performance is very close to the minimum outage
probability obtained by optimal schemes. We also show the
applicability of these techniques to practical systems employing
orthogonal frequency division multiplexing.
\end{abstract}

\newpage


\section{Introduction}

A key design challenge for wireless communications systems is to
provide high-data-rate wireless access, while optimizing the use
of limited resources such as available frequency bandwidth,
transmission power and computational ability of portable devices.
Reliable transmission is particular challenging for wireless
communications systems due to the harsh, time-varying signal
propagation environment. Mobility and multipath propagation
\cite{BiglieriProakisShamai1998,BenedettoBiglieri1999,Biglieri2006} lead to
time-selective and frequency selective fading channels, where the
dynamics of the signal variations depend on mobile velocity,
carrier frequency, transmission bandwidth, and the particular
scattering environment.

The use of {\em orthogonal frequency division multiplexing} (OFDM)
technologies is a proven approach for providing high data rates in
wireless communications systems. Standards such as IEEE 802.11
(WiFi) \cite{NandaWaltonKetchumWallaceHoward2005} and IEEE 802.16
(WiMax) \cite{GhoshWoltersAndrewsChen2005} already include OFDM as
a core technology, and future generations of mobile cellular
systems are likely to also feature multi-carrier techniques. OFDM
transmission over frequency-selective or time-frequency-selective
wireless fading channels is adequately modelled as a {\em
block-fading channel}.

The block-fading channel \cite{OzarowShamaiWyner1994,
BiglieriProakisShamai1998} is a useful channel model for a class
of time- and/or frequency-varying fading channels where the
duration of a block-fading period is determined by the product of
the channel coherence bandwidth and the channel coherence time
\cite{Proakis1995}. Within a block-fading period, the channel
fading gain remains constant, while between periods the channel
gains change according to a system-specific rule. In this setting,
transmission typically extends over multiple block-fading periods.
Frequency-hopping schemes as encountered in the Global System for
Mobile Communication (GSM) and the Enhanced Data GSM Environment
(EDGE), as well as transmission schemes based on multiple antenna
systems, can also conveniently be modelled as block-fading
channels. The simplified model is mathematically tractable, while
still capturing the essential features of the practical
transmission schemes over fading channels.

In many situations of practical interest, channel state
information (CSI), namely the degree of knowledge that either the
transmitter, the receiver, or both, have about the channel gains,
greatly influences system design and performance. In general, optimal
transmission strategies over a block-fading channel depend on the
availability of CSI at both sides of the transmission link
\cite{BiglieriProakisShamai1998}. At the receiver side,
time-varying channel parameters can often be accurately estimated
\cite{Proakis1995}. Thus, perfect CSI at the receiver (CSIR) is a
common and reasonable assumption. Conversely, perfect CSI at the transmitter
(CSIT) depends on the specific system architecture. In a system with
time-division duplex (TDD), the same channel can be used for both
transmission and reception, provided that the channel varies slowly. In this case, perfect CSIR can be used
reciprocally as perfect CSIT \cite{KnoppCaire2002}. In other
system architectures, CSIT is provided through
channel-state-feedback from the receiver.
When no CSIT is available, transmit power is commonly allocated
uniformly over the blocks. In contrast, when CSIT is available,
the transmitter can adapt the transmission mode (transmission
power, data rate, modulation and coding) to the instantaneous
channel characteristics, leading to significant performance
improvements.

We distinguish between two cases of transmission dynamics. On the one hand, if no delay constraints are enforced, transmission
extends over a large (infinite) number of fading blocks. The
corresponding fading process is stationary and ergodic, revealing
the fading statistics during the transmission. The maximum data
rate for this case, termed the ergodic capacity, was determined in
\cite{GoldsmithVaraiya1997}, assuming perfect CSI at both
transmitter and receiver. Two coding schemes have been shown to
achieve the ergodic capacity. In \cite{GoldsmithVaraiya1997}, a
variable-rate, variable-power transmission strategy based on a
library of codebooks, and driven by the CSIT, was suggested. In
contrast, a fixed-rate, variable-power transmission strategy was
proposed in \cite{CaireShamai1999} based on a single codebook,
providing a practically more appealing alternative in the form of
a conventional Gaussian encoder followed by power allocation
driven by the CSIT.

On the other hand, when a delay constraint is enforced, the
transmission of a codeword only spans a finite number of fading
blocks. This constraint corresponds to real-time transmission over
slowly varying channels. Therefore, this situation is relevant for
wireless OFDM applications in wireless local area networks (WLAN).
As the channel relies on particular realizations of the finite
number of independent fading coefficients, the channel is
non-ergodic and therefore not information stable
\cite{VerduHan1994,CaireTariccoBiglieri1999}. It follows that the
Shannon capacity under most common fading statistics is zero,
since there is an irreducible probability, denoted as the {\em
outage probability}, that the channel is unable to support the
actual data rate,
\cite{OzarowShamaiWyner1994,BiglieriProakisShamai1998}. For
sufficiently long codes, the word error rate is strictly
lower-bounded by the outage probability. In some cases there is
a maximum non-zero rate and a minimum finite signal-to-noise ratio
(SNR) for which the minimum outage probability is zero. This
maximum rate is commonly referred to as the {\em delay-limited
capacity} \cite{HanlyTse1998}. In this paper, we will consider
fixed-rate transmission strategies over delay-limited non-ergodic
block-fading channels.

In a practical system, only causal CSIT is available. Thus, in
general, the channel gains are only known up to (and possibly
including) the current block-fading period. However, in an OFDM
system with multiple parallel carriers, the causal constraint
still allows for all sub-carrier channel gains to be known
simultaneously in a seemingly non-causal manner, as compared to a
block-fading channel based on frequency-hopping single-carrier
transmission. Here, we will only consider the OFDM-inspired
scenario where perfect non-causal CSIT is available.

As mentioned above, when perfect CSI is available at the
transmitter, power allocation techniques can be used to increase
the instantaneous mutual information, thus improving the outage
performance. Multiple power allocation rules derived under a
variety of constraints have been proposed in the literature
\cite{CaireTariccoBiglieri1999, DeyEvans2005, DeyEvans2007,
LuoLinYatesSpasojevic2003, LuoYatesSpasojevic2005}. The optimal
power allocation minimizes the outage probability subject to a
{\em short-term} power constraint over a single codeword or a {\em
long-term} power constraint over all transmitted codewords. The
optimal transmission strategy, subject to a short-term power
constraint, was shown in \cite{CaireTariccoBiglieri1999} to
consist of a random code with independent, identically distributed
Gaussian code symbols, followed by optimal power allocation based
on water-filling \cite{CoverThomas2006}. The optimal power
allocation problem is also solved in
\cite{CaireTariccoBiglieri1999} under a long-term power
constraint, showing that remarkable gains are possible with
respect to transmission schemes with short-term power constraints.
In some cases, the optimal power-allocation scheme can even
eliminate outages, leading to a minimum outage probability
approaching zero
\cite{CaireTariccoBiglieri1999,BiglieriCaireTaricco2001}, and
thus, a non-zero delay-limited capacity. In particular, gains of
more than $12$ dB are possible at practically relevant error
probabilities. Again, the optimal input distribution is Gaussian.
The optimal power allocation problem under a long-term power
constraint, and with perfect CSIR but only partial CSI available
at the transmitter is considered in \cite{KimSkoglund2007}. The
problem is solved for the limiting case of large SNR, leading to
similar impressive improvements in outage performance.

In practical wireless communications systems, coding schemes are
constructed  over discrete signal constellations, e.g., PSK, QAM.
It is therefore of practical interest to derive power allocation
rules for coded modulation schemes with discrete input
constellations, minimizing the outage probability. A significant
step towards this goal was achieved in
\cite{LozanoTulinoVerdu2006}, where the fundamental relationship
between mutual information and MMSE developed in
\cite{GuoShamaiVerdu2005} proved instrumental to optimizing the
transmit power of parallel channels with discrete inputs. As
stated in \cite{LozanoTulinoVerdu2006}, the developed power
allocation rule for parallel channels can be applied directly to
minimize the outage probability of delay-limited block-fading
channels under short-term power constraints. However, the optimal
solution in \cite{LozanoTulinoVerdu2006} does not reveal the
impact of the system parameters involved, and may also be
prohibitively complex for practical applications with
computational power and memory limitations.

In this paper, we study power allocation schemes that minimize the
outage probability of {\em fixed-rate} coded modulation schemes
using discrete signal constellations under short- and long-term
power constraints. In particular, we study classical coded modulation schemes, as well as bit-interleaved coded modulation (BICM) using suboptimal non-iterative decoding \cite{CaireTariccoBiglieri1998}.
Similarly to the uniform power allocation case, we show
that under a short-term power constraint, an application of the
Singleton bound \cite{MalkamakiLeib1999,KnoppHumblet2000,FabregasCaire2006,
NguyenGuillenRasmussen2006}
leads to the optimal SNR exponent (diversity gain) of the channel. In particular, we show that for Nakagami-$m$ channels, the optimal SNR exponent is given by $m$ times the Singleton bound \cite{MalkamakiLeib1999,KnoppHumblet2000,FabregasCaire2006,
NguyenGuillenRasmussen2006}.
In the long-term case, we derive the optimal power allocation
scheme. We show that the underlying structure of the solution for
Gaussian inputs in \cite{CaireTariccoBiglieri1999} remains valid,
where no power is allocated to bad fading realizations, minimizing
power wastage. We also show that the relationship between the
mutual information and the minimum-mean-squared-error (MMSE)
reported in \cite{GuoShamaiVerdu2005} is instrumental in deriving
the optimal outage-minimizing long-term solution\footnote{The
optimal power allocation algorithm with discrete inputs has been
independently reported in \cite{CaireKumar2007}. We became aware
of the results in \cite{CaireKumar2007} after the submission of
the conference version of this paper
\cite{NguyenGuillenRasmussen2007itw}.}. We analyze the optimal
long-term solution, showing that zero outage can be achieved
provided that the SNR exponent corresponding to the short-term scheme with the same parameters is strictly greater than one,
implying the delay-limited capacity is non-zero. Conversely, if
the short-term SNR exponent is smaller than one, we show that zero outage
cannot be achieved. In this case, we derive the corresponding long-term SNR
exponent as a function of the Singleton bound.

Practical transmitters may have limited memory and computational
resources that may prevent the use of the optimal solution based
on the MMSE. We further aim at reducing the computational
complexity and memory requirements of optimal schemes by proposing
sub-optimal short- and long-term power allocation schemes. The
sub-optimal schemes enjoy significant reductions in complexity,
yet they only suffer marginal performance losses as compared to
relevant optimal schemes. For the suggested sub-optimal schemes,
we further characterize the corresponding SNR exponents and
delay-limited capacities for short- and long-term constraints as
functions of the modified Singleton bound.

The paper is further organized as follows. In Section
\ref{se:model}, the system model, basic assumptions and related
notation are described, while mutual information, MMSE and outage
probability are introduced in Section \ref{se:info_theory}.
Optimal and sub-optimal power allocation schemes with short-term
power constraints are considered in Section \ref{se:short_term},
and corresponding optimal and sub-optimal power allocation schemes
with long-term power constraints are investigated in Section
\ref{se:long_term}. Numerical examples are used throughout the
paper to illustrate the presented results. Concluding remarks,
summarized in Section \ref{se:conclude}, complete the main body of
the paper. To support the readability of the paper, lengthy proofs
are moved to appendices.

Throughout the paper, we shall make use of the following notation.
Scalar and vector variables are characterized with lowercase and
boldfaced lowercase letters, respectively. The expectation with
respect to the fading statistics is simply denoted by
$\expectation{\cdot}{}$, while the expectation with respect to any
other arbitrary random variable $\Phi$ is denoted by
$\expectation{\cdot}{\Phi}$. Furthermore, the expectation with
respect to an arbitrary random variable $\Phi$, with the
constraint $\Phi \in \mathcal{R}$ is denoted as
$\expectation{\cdot}{\Phi \in \mathcal{R}}$. We define
$\mean{\mathbf{x}} \triangleq \frac{1}{B} \sum_{i=1}^B x_i$ as the
arithmetic mean of $\mathbf{x}=(x_1, x_2,...,x_B)$. The
exponential equality $f(\xi) \doteq K\xi^{-d}$ indicates that
$\lim_{\xi \to \infty}f(\xi) \xi^d = K $, with the exponential
inequalities $\dot{\leq},\dot{\geq}$ similarly defined.
$\mathbb{R}^{n}_{+}= \left\{\xi \in \mathbb{R}^{n}|\xi> 0
\right\}$, $\min\{a,b\}$ denotes the minimum of $a$ and $b$,
$\lceil \xi \rceil$ $\left(\lfloor \xi \rfloor\right)$ denotes the
smallest (largest) integer greater (smaller) than $\xi$, while
$(f(x))_+ = 0$ if $f(x)<0$, and $(f(x))_+ = f(x)$ if $f(x)\geq 0$.

\section{System Model}
\label{se:model}

\begin{figure}
\begin{center}
\includegraphics[width=0.8\columnwidth] {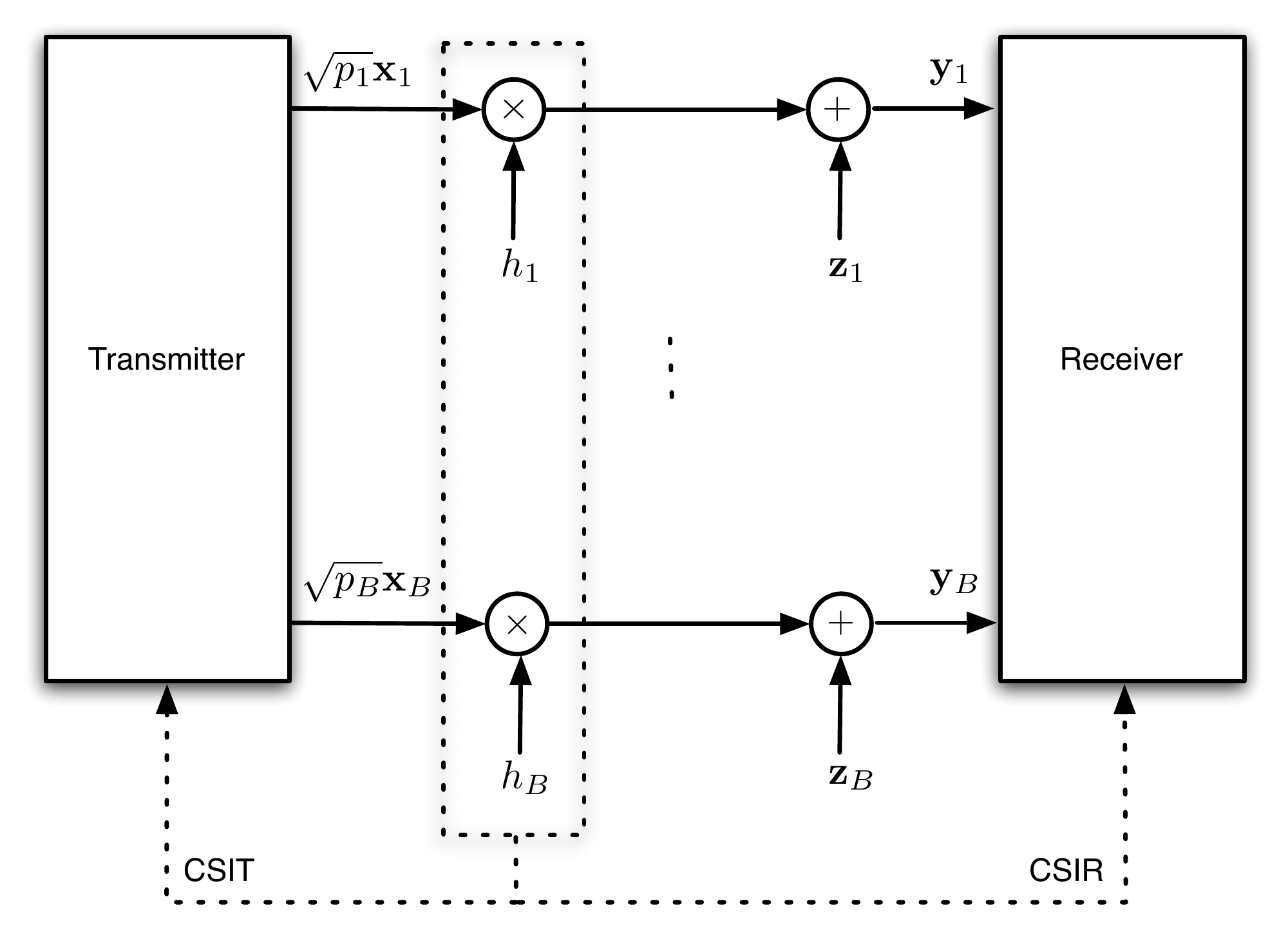}
\caption{Block diagram corresponding to the channel system model with CSI at the transmitter and the receiver.}
\label{fig:model}
\end{center}
\end{figure}

Consider transmission over an additive white Gaussian noise (AWGN)
block-fading channel with $B$ blocks of $L$ channel uses each, in
which, for $b=1, \ldots, B$, block $b$ is affected by a flat
fading coefficient $h_b\in \mathbb{C}$. The corresponding block
diagram is shown in Figure \ref{fig:model}. Let $\gamma_b=
|h_b|^2$ be the power fading gain and assume that the fading gain
vector $\boldgamma = (\gamma_1, \ldots, \gamma_B)$ is available at
both the transmitter and the receiver. The transmit power is
allocated to the blocks according to the scheme
$\mathbf{p}(\boldgamma)=(p_1(\boldgamma), \ldots,
p_B(\boldgamma))$. Then, the complex baseband channel model can be
written as
\begin{equation}
\label{eq:channel_model} \mathbf{y}_b = \sqrt{p_b(\boldgamma)}h_b \mathbf{x}_b
+ \mathbf{z}_b, \ \ b= 1, \ldots, B,
\end{equation}
where $\mathbf{y}_b \in \mathbb{C}^L$ is the received signal in block $b$,
$\mathbf{x}_b \in \mathcal{X}^L \subset \mathbb{C}^L$ is the portion of the
codeword being transmitted in block $b$, $\mathcal{X} \subset \mathbb{C}$ is
the signal constellation and $\mathbf{z}_b \in \mathbb{C}^L$ is a noise vector
with independent, identically distributed (i.i.d.) circularly symmetric
Gaussian entries $\sim \mathcal{N}_{\mathbb{C}}(0, 1)$. Assume that the signal
constellation $\mathcal{X}$ is normalized in energy such that
$\sum_{x\in\mathcal{X}}|x|^2 = 2^M$, where $M=\log_2|\mathcal{X}|$. Then, the
instantaneous received SNR at block $b$ is given by $p_b(\boldgamma) \gamma_b$.
The following power constraints are considered \cite{CaireTariccoBiglieri1999}
\begin{align}
\text{Short-Term:}&\;\;\;\;\;  \mean{\mathbf{p}(\boldgamma)} \triangleq \frac{1}{B}
\sum_{b=1}^B p_b(\boldgamma) \leq P\\
\text{Long-Term:}& \;\;\;\;\; \expectation{\mean{\mathbf{p}(\boldgamma)}}{} =
\expectation{\frac{1}{B} \sum_{b=1}^B p_b(\boldgamma)}{} \leq P.
\end{align}
In all cases, $P$ represents the average SNR at the receiver. We
will denote by $\pst$ and $\plt$ the power allocation vectors
corresponding to short- and long-term power constraints,
respectively. We will also denote by $\peq(p)=(p,\dotsc,p)$ the
uniform power vector that allocates the same power $p$ to each
block.

We consider block-fading channels where $h_b$ are independent realizations of a
random variable $H$, whose magnitude is Nakagami-$m$ distributed
\cite{Nakagami1960, SimonAlouini2005} and has a uniformly distributed
phase\footnote{Due to our perfect transmitter and receiver CSI assumption, we
can assume that the phase has been perfectly compensated for.}. The fading
magnitude has the following probability density function (pdf)
\begin{equation}
f_{|H|}(h)= \frac{2m^m h^{2m-1}}{\Gamma(m)}e^{-m h^2},
\end{equation}
where $\Gamma(a)$ is the Gamma function, $\Gamma(a) =
\int_0^{\infty}t^{a-1}e^{-t}dt$ \cite{abramowitzStegun1964}. The
coefficients $\gamma_b$ are realizations of the random variable
$|H|^2$ whose pdf and cdf are given by
\begin{equation}
\label{eq:sq_Naka_dist}
f_{\gamma_b}(\gamma)= \begin{cases}
\frac{m^m \gamma^{m-1}}{\Gamma(m)}e^{-m\gamma}, & \gamma \geq 0\\
0, &{\rm otherwise}
\end{cases}
\end{equation}
and
\begin{equation}
F_{\gamma_b}(\xi)= \begin{cases}
1 - \frac{\Gamma(m, m \xi)}{\Gamma(m)}, &\xi \geq 0\\
0, &{\rm otherwise},
\end{cases}
\end{equation}
respectively, where $\Gamma(a, \xi)$ is the upper incomplete Gamma function,
$\Gamma(a, \xi)= \int_{\xi}^{\infty} t^{a-1}e^{-t}dt$ \cite{abramowitzStegun1964}.

The Nakagami-$m$ distribution encompasses many fading distributions of
interest. In particular, we obtain Rayleigh fading by letting $m=1$ and an
accurate approximation of Rician fading with parameter $K$ by setting
$m=(K+1)^2/(2K+1)$ \cite{SimonAlouini2005}.

\section{Mutual Information, MMSE and Outage Probability}
\label{se:info_theory}

The channel model described in \eqref{eq:channel_model}
corresponds to a parallel  channel model, where each sub-channel
is used a fraction $\frac{1}{B}$ of the total number of channel
uses per codeword. Therefore, for any given power fading gain
realization $\boldgamma$ and power allocation scheme
$\mathbf{p}(\boldgamma)$, the instantaneous input-output mutual
information of the channel is given by \cite{CoverThomas2006}
\begin{equation}
I_B(\mathbf{p}(\boldgamma), \boldgamma) = \frac{1}{B}\sum_{b=1}^B \IX(p_b \gamma_b),
\end{equation}
where $\IX(\rho)$ is the input-output mutual information of an
AWGN channel with input constellation ${\mathcal X}$ and received
SNR $\rho$. In this paper we will consider coded modulation (CM)
schemes with uniform  input distribution, for which  $\IX(\rho)$
is given by
\begin{equation}
  \IX^{\rm CM}(\rho) = M - \frac{1}{2^M}\sum_{x\in\mathcal{X}}
  \mathbb{E}_{Z}\left[\log_2\left(\sum_{x'  \in  \mathcal X}
  e^{-|\sqrt{\rho}(x-x')+Z|^2+|Z|^2}\right)\right].\nonumber
\end{equation}
Furthermore, we also consider bit-interleaved coded modulation
(BICM)  using the classical sub-optimal non-iterative BICM decoder
proposed in \cite{Zehavi1992}, for which the mutual information
for a given binary labeling rule\footnote{We select Gray labeling
\cite{Proakis2001}, since it has been shown to maximize the mutual
information for the non-iterative BICM decoder
\cite{CaireTariccoBiglieri1998}.} can be expressed as the mutual
information of $M$ binary-input continuous-output-symmetric
parallel channels \cite{CaireTariccoBiglieri1998},
\begin{align}
  \IX^{\rm BICM}(\rho)
  &= M -  \frac{1}{2^M}\sum_{q=0}^1\sum_{j=1}^{M}\sum_{x\in\mathcal{X}_q^j}
  \mathbb{E}_{Z}\left[\log_2\frac{ \displaystyle \sum_{x'  \in  \mathcal  X}
  e^{-|\sqrt{\rho}(X-x')+Z|^2}}{ \displaystyle \sum_{x'  \in  \mathcal{X}_q^j}
  e^{-|\sqrt{\rho}(X-x')+Z|^2}}\right].\nonumber
\end{align}
where the sets $\mathcal{X}_c^i$ contain all signal constellation
points with bit $c$ in the $j$-th binary labeling position. Both
CM and BICM mutual information expressions can be efficiently
evaluated numerically using Gauss-Hermite quadratures
\cite{abramowitzStegun1964}.

A fundamental relationship between the MMSE and the mutual
information (in bits) in additive Gaussian channels is introduced
in \cite{GuoShamaiVerdu2005} showing that
\begin{equation}
\frac{d}{d\rho} \IX(\rho)= \frac{1}{\log 2}\MMSE(\rho),
\label{eq:mmse_i}
\end{equation}
where $\MMSE(\rho)$ is the MMSE for a given signal constellation
$\mathcal{X}$ expressed as a function of the SNR $\rho$. This
relationship proves instrumental in obtaining optimal power control
rules. In particular, for the CM case, the MMSE resulting from
estimating the input based on the received signal over an AWGN
channel with SNR $\rho$ can be written as
\cite{LozanoTulinoVerdu2006},
\begin{align}
\MMSE^{\rm CM}(\rho) &= \expectation{|X-\hat{X}|^2}{}=\expectation{|X-\expectation{X|Y}{}|^2}{}\\
&=1 - \frac{1}{\pi} \int_{\mathbb{C}} \frac{\left|\sum_{x\in
\mathcal{X}} x e^{-|y-\sqrt{\rho}x|^2}\right|^2}{\sum_{x \in
\mathcal{X}}e^{-|y-\sqrt{\rho}x|^2}} dy, \label{eq:mmse}
\end{align}
where $\hat{X}=\expectation{X|Y}{}$ is the MMSE estimate of $X$
given the channel observation $Y$. Once again, \eqref{eq:mmse} can
be efficiently evaluated using Gauss-Hermite quadratures
\cite{abramowitzStegun1964}. In the case of BICM, we obtain an
equivalent set of symmetric binary-input continuous output channels
\cite{CaireTariccoBiglieri1998}. However, due to the demodulation
process, the equivalent channels have a noise that is non-Gaussian,
and more importantly, non-additive. We therefore {\em define} the
function derivative of $\IX^{\rm BICM}(\rho)$, denoted by
$\MMSE^{\rm BICM}(\rho)$, by enforcing the relationship
\eqref{eq:mmse_i} to hold as follows
 \begin{equation}
\MMSE^{\rm BICM}(\rho) \triangleq \frac{d}{d\rho} \IX^{\rm BICM}(\rho).
\label{eq:mmse_bicm}
\end{equation}
Note that this is only a shorthand notation, so that whenever
$\MMSE$  appears in the coming sections, it can be replaced by
either $\MMSE^{\rm CM}(\rho)$ or $\MMSE^{\rm BICM}(\rho)$.
Therefore, \eqref{eq:mmse_bicm} {\em does not} denote the MMSE in
estimating the input bits given the noisy channel observation. The
function  $\MMSE^{\rm BICM}(\rho)$ can again be easily evaluated
numerically.

Finally, we define the transmission to be in outage when the
instantaneous input-output mutual information is less than the
target {\em fixed} transmission rate $R$. For a given power allocation scheme
$\mathbf{p}(\boldgamma)$ with power constraint $P$, the outage
probability at transmission rate $R$ is given by
\cite{OzarowShamaiWyner1994,BiglieriProakisShamai1998}
\begin{align}
\label{eq:out_prob}
\Pout(\mathbf{p}(\boldgamma), P, R) &=\Pr(I_B(\mathbf{p}(\boldgamma), \boldgamma)< R) \nonumber\\
&=\Pr\left(\frac{1}{B} \sum_{b=1}^B \IX(p_b\gamma_b)< R\right).
\end{align}
Since  $ \IX^{\rm BICM}(\rho)\leq \IX^{\rm CM}(\rho)$ we will have
that the corresponding outage probabilities verify that
$\Pout^{\rm BICM}(\mathbf{p}(\boldgamma), P, R) \geq \Pout^{\rm
CM}(\mathbf{p}(\boldgamma), P, R)$. All the algorithms and results
presented in the following are valid for both CM and BICM.
Therefore, unless explicitly stated, we will use the common
notation $\IX(\rho)$ and $\MMSE(\rho)$ to refer to both.

\section{Short-Term Power Allocation}
\label{se:short_term}

Short-term power allocation schemes are applied to systems where
the transmit power of each codeword is limited to $BP$. Following
the definition of short-term power constraint given in Section
\ref{se:model}, a given short-term power allocation scheme
$\pst=(p_1, \ldots, p_B)$ must then satisfy
$\frac{1}{B}\sum_{b=1}^B p_b \leq P$.

\subsection{Optimal Short-Term Power Allocation}
The optimal short-term power allocation rule $\mathbf{p}^{\rm opt}_{\rm st}(\boldgamma)$
is the solution to the outage probability minimization problem
\cite{CaireTariccoBiglieri1999}. Mathematically we express $\mathbf{p}^{\rm
opt}_{\rm st}(\boldgamma)$ as
\begin{equation}
\label{eq:opt_short_prob} \mathbf{p}^{\rm opt}_{\rm st}(\boldgamma) = \arg
\min_{\substack{{\mathbf p}\in \mathbb{R}^B_+\\ \frac{1}{B}\sum_{b=1}^B p_b = P} }
\Pout(\mathbf{p},P, R).
\end{equation}

For short-term power allocation, the power allocation scheme that maximizes the
instantaneous mutual information at each channel realization also minimizes the
outage probability since the available power can only be
distributed within one codeword. Formally, we have \cite{CaireTariccoBiglieri1999}
\begin{lemma}
\label{le:min_pout_sol} Let $\mathbf{p}^{\rm opt}_{\rm st}(\boldgamma)$ be a solution of
the problem
\begin{equation}
\label{eq:prob_max_cap}
\begin{cases}
\text{Maximize} & \sum_{b=1}^B \IX(p_b \gamma_b)\\
\text{Subject to}& \frac{1}{B}\sum_{b=1}^B p_b  \leq P\\
&p_b \geq 0, b=1, \ldots, B.
\end{cases}
\end{equation}
Then $\mathbf{p}^{\rm opt}_{\rm st}(\boldgamma)$ is a solution of $\eqref{eq:opt_short_prob}$.
\end{lemma}
\begin{proof}
See Appendix \ref{app:short_opt_proof}.
\end{proof}

The solution of problem \eqref{eq:prob_max_cap}, which is based on
the relationship between the MMSE and the mutual information
\cite{GuoShamaiVerdu2005}, was obtained in
\cite{LozanoTulinoVerdu2006}. From \cite{LozanoTulinoVerdu2006}
one has the following theorem.
\begin{theorem}
\label{theorem:optimal_st} The solution of problem \eqref{eq:prob_max_cap} is
given by
\begin{equation}
\label{eq:opt_sol_short}
\popt_b(\boldgamma) =
\frac{1}{\gamma_b}\MMSE^{-1}\left(\min\left\{1,
\frac{\eta}{\gamma_b}\right\}\right),
\end{equation}
for $b=1, \dotsc, B$, where $\eta$ is chosen such that the power
constraint is satisfied,
\begin{equation}
\frac{1}{B}\sum_{b=1}^B \frac{1}{\gamma_b}\MMSE^{-1}\left(\min\left\{1,
\frac{\eta}{\gamma_b}\right\}\right) = P.
\end{equation}
\end{theorem}

\begin{proof}
See \cite{LozanoTulinoVerdu2006} for details.
\end{proof}
As outlined in \cite{LozanoTulinoVerdu2006}, the results of Theorem
\ref{theorem:optimal_st}, are valid for any constellation. In particular, since
the MMSE for Gaussian inputs (GI) is
\begin{equation}
\MMSE^{\rm GI}(\rho) = \frac{1}{\log 2 \, (1+\rho)}, \label{eq:mmse_gi}
\end{equation}
the inverse function can be written in closed form \cite{LozanoTulinoVerdu2006}
and we therefore recover the waterfillng solution by replacing $\MMSE(\rho)$ by
$\MMSE^{\rm GI}(\rho)$ \cite{CaireTariccoBiglieri1999}.

The optimal short-term power allocation scheme improves the outage performance
of coded modulation schemes over block-fading channels. However, it does not
increase the outage diversity compared to a uniform power allocation, as shown
in the following result.
\begin{proposition}
\label{le:opt_diver} Consider transmission over the block-fading channel
defined in \eqref{eq:channel_model} with the optimal power allocation scheme
$\pst^{\rm opt}(\boldgamma)$ given in \eqref{eq:opt_sol_short}. Assume
input constellation size $|\mathcal{X}|= 2^M$. Further assume that the power
fading gains follow the distribution given in \eqref{eq:sq_Naka_dist}. Then,
for large $P$ and some $ \mathcal{K}_{\rm opt}>0$ the outage probability
behaves as
\begin{equation}
\Pout(\pst^{\rm opt}(\boldgamma), P,R) \doteq \mathcal{K}_{\rm opt}P^{-m d_B(R)},
\end{equation}
where $d_B(R)$ is the Singleton bound given by
\begin{equation}
d_B(R)= 1+ \left\lfloor B\left(1-\frac{R}{M}\right)\right\rfloor.
\label{eq:sb}
\end{equation}
\end{proposition}
\begin{proof}
See Appendix \ref{app:short_opt_proof}.
\end{proof}

\subsection{Sub-optimal Short-Term Power Allocation Schemes}
Although the power allocation scheme in \eqref{eq:opt_sol_short}
is optimal, it involves an inverse MMSE function, which may be
excessively complex to implement or store for specific low-cost
systems. Moreover, the MMSE function provides little insight into
the role of each system parameter. In this section, we propose
sub-optimal power allocation schemes similar to water-filling that
tackle both drawbacks, leading to only minor losses in outage
performance as compared to the optimal solution.

\subsubsection{Truncated water-filling scheme}
\label{se:wlikesol} 
\begin{figure}
\begin{center}
\includegraphics[width  =1 \columnwidth] {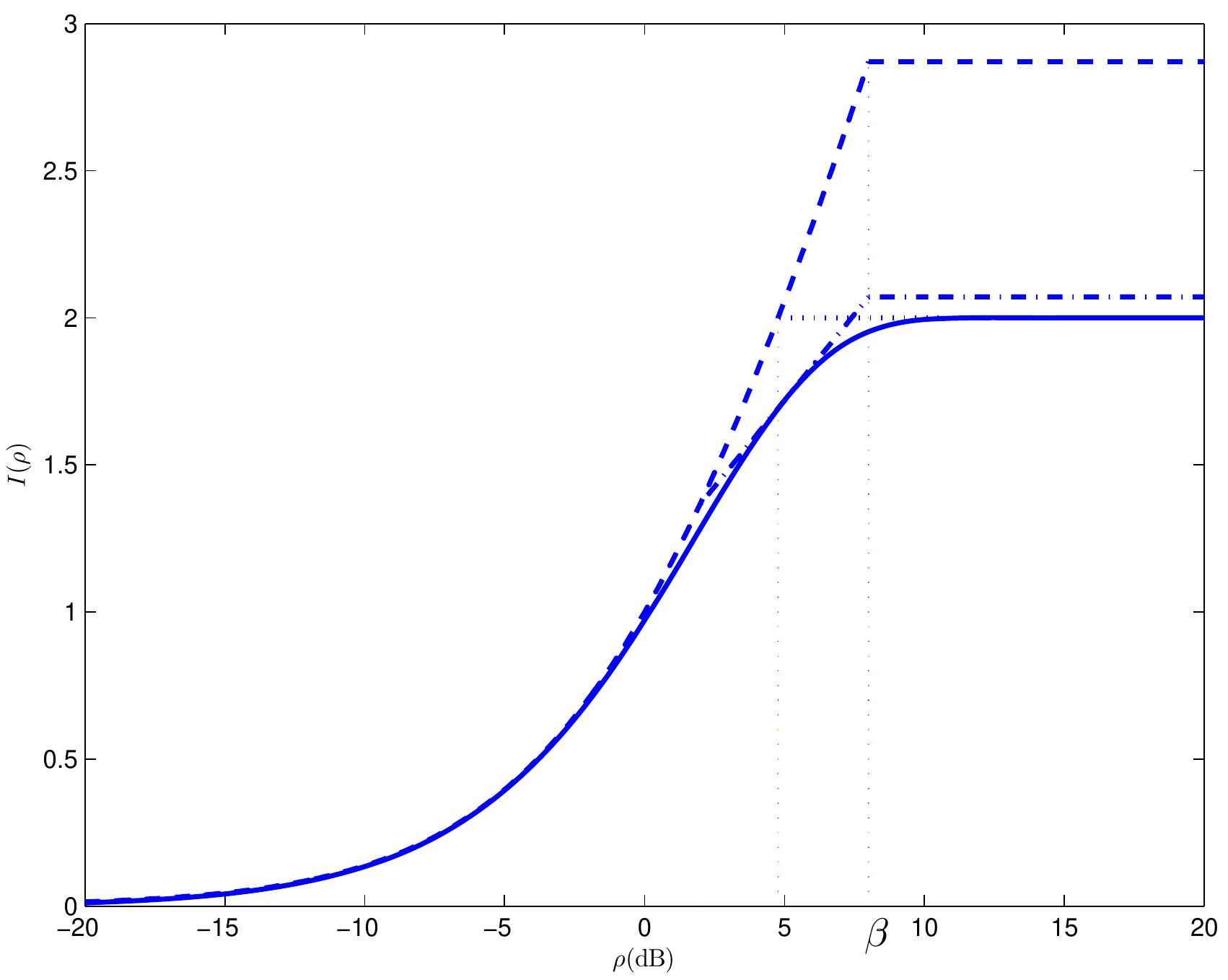}
\caption{Mutual information of QPSK, and the approximations used
by truncated water-filling and its corresponding refinement.}
\label{fig:refine_bound_extended}
\end{center}
\end{figure}

The complexity of the solution in
\eqref{eq:opt_sol_short} is due to the complex expression of
$\IX(\rho)$ in problem \eqref{eq:prob_max_cap}. Therefore, in
order to obtain a simple sub-optimal solution, we propose an
approximation for $\IX(\rho)$ in problem \eqref{eq:prob_max_cap}.
For Gaussian input channels with $I(\rho)= \log_2(1+\rho)$,
optimal power allocation is obtained by the simple water-filling
scheme \cite{CoverThomas2006}. This suggests the use of the
following approximation for $\IX(\rho)$.
\begin{equation}
\label{eq:target_tw}
\IX(\rho) \leq I^{\rm tw}(\rho) \triangleq \begin{cases} \log_2(1+\rho), &
  \rho \leq \beta\\
\log_2(1+\beta), &\text{otherwise,}
\end{cases}
\end{equation}
where $\beta$ is a design parameter to be optimized for best performance. An
example of $I^{\rm tw}(\rho)$ is illustrated in Figure
\ref{fig:refine_bound_extended}. The resulting sub-optimal scheme
$\mathbf{p}_{\rm st}^{\rm tw}(\boldgamma)$ is given as a solution of
\begin{equation}
\label{eq:trunc_wfill_prob}
\begin{cases}\text{Maximize} &\sum_{b=1}^B I^{\rm tw}(p_b \gamma_b)\\
\text{Subject to} &\frac{1}{B} \sum_{b=1}^B p_b \leq P\\
&p_b \geq 0, \ b=1, \ldots, B.
\end{cases}
\end{equation}

\begin{theorem}
\label{le:trunc_wfill_sol} A solution to the problem in
\eqref{eq:trunc_wfill_prob} is given by
\begin{equation}
\label{eq:trunc_wfill_sol}
\ptwfill_b (\boldgamma)= \begin{cases}
\frac{\beta}{\gamma_b}, & {\rm if} ~\frac{1}{B}\sum_{b=1}^B \frac{\beta}{\gamma_b} \leq
P\\
\min\left\{\frac{\beta}{\gamma_b}, \left(\eta -
\frac{1}{\gamma_b}\right)_+\right\}, &\text{otherwise},
\end{cases}
\end{equation}
for $b=1, \ldots, B$, where $\eta$ is chosen such that
\begin{equation}
\frac{1}{B}\sum_{b=1}^B \min\left\{\frac{\beta}{\gamma_b}, \left(\eta -
\frac{1}{\gamma_b}\right)_+\right\} = P.
\end{equation}
\end{theorem}
\begin{proof}
See Appendix \ref{app:tw_sol}.
\end{proof}

Without loss of generality, assume that $\gamma_1 \geq \ldots \geq \gamma_B$,
then, similarly to water-filling, $\eta$ can be determined such that \cite{CaireTariccoBiglieri1999}
\begin{equation}
(k-l)\eta = BP - \sum_{b=1}^l \frac{\beta+1}{\gamma_b}+
\sum_{b=1}^k \frac{1}{\gamma_b},
\end{equation}
where $k, l$ are integers satisfying $\frac{1}{\gamma_k}< \eta <
\frac{1}{\gamma_{k+1}}$ and $\frac{\beta+1}{\gamma_l}< \eta \leq
\frac{\beta+1}{\gamma_{l+1}}$.

From  Theorem \ref{le:trunc_wfill_sol}, the resulting power
allocation scheme is similar to water-filling, except for the
truncation of the allocated power at $\frac{\beta}{\gamma_b}$. We
refer to this scheme as truncated water-filling. The outage
performance obtained by the truncated water-filling scheme depends
on the choice of the design parameter $\beta$. We now analyze the
asymptotic performance of the outage probability, thus providing
some guidance for the choice of $\beta$.

\begin{proposition}
\label{le:tw_diver} Consider transmission over the block-fading
channel defined in \eqref{eq:channel_model} with input
constellation $\mathcal{X}$ and the truncated water-filling power
allocation scheme $\mathbf{p}^{\rm tw}_{\rm st}(\boldgamma)$ given
in \eqref{eq:trunc_wfill_sol}. Assume that the power fading gains
follow the distribution given in \eqref{eq:sq_Naka_dist}. Then,
for large $P$, the outage probability $\Pout(\mathbf{p}^{\rm
tw}_{\rm st}(\boldgamma),P, R)$ is asymptotically upper bounded by
\begin{equation}
\label{eq:trunc_wfill_outprob_bound}
\Pout(\mathbf{p}^{\rm tw}_{\rm st}(\boldgamma),P, R)~\dot{\leq}~
\mathcal{K}_{\beta} P^{-m  d_{\beta}(R)},
\end{equation}
where
\begin{equation}
d_{\beta}(R) =
1+\left\lfloor B\left(1-\frac{R}{I_{\mathcal X}(\beta)}\right)\right\rfloor,
\end{equation}
and $I_{\mathcal{X}}(\beta)$ is the input-output mutual
information of an AWGN channel with SNR $\beta$.
\end{proposition}
\begin{proof}
See Appendix \ref{app:proof_tw_dive}.
\end{proof}

From the results of Proposition \ref{le:opt_diver}, Proposition
\ref{le:tw_diver}, and noting that
$\Pout(\mathbf{p}^{\rm tw}_{\rm st}(\boldgamma),P, R) \geq
\Pout(\mathbf{p}^{\rm opt}_{\rm st}(\boldgamma),P, R)$, we have
\begin{equation}
\Pout(\mathbf{p}^{\rm tw}_{\rm st}(\boldgamma),P, R) \doteq \mathcal{K}_{\rm tw} P^{-md_{\rm
    tw}(R)},
\end{equation}
where $d_{\rm tw}(R)$ satisfies that $d_\beta(R) \leq d_{\rm tw}(R) \leq
d_B(R)$. Therefore, the truncated water-filling scheme is guaranteed to obtain optimal
diversity whenever $d_\beta(R)= d_B(R)$, or equivalently, when
\begin{align}
B\left(1-\frac{R}{I_{\mathcal X}(\beta)}\right) &\geq
\left\lfloor B\left(1-\frac{R}{M}\right)\right\rfloor\\
I_{\mathcal X}(\beta) &\geq \frac{BR}{B- \left\lfloor B
  \left(1-\frac{R}{M}\right)\right\rfloor},
  \label{eq:boundbeta}
\end{align}
which implies that
\[
\beta \geq I_{\mathcal X}^{-1} \left(\frac{BR}{B- \left\lfloor B
\left(1-\frac{R}{M}\right)\right\rfloor}\right) \triangleq \beta_R.
\]
Therefore, by letting $\beta \to \infty$, the truncated
water-filling power allocation scheme given in
\eqref{eq:trunc_wfill_sol}, which now  becomes the classical
water-filling algorithm for Gaussian inputs, provides optimal
outage diversity at any transmission rate. For any rate $R$ that
is not at a discontinuity point of the Singleton bound, i.e. $R$
such that $B\left(1-\frac{R}{M}\right)$ is not an integer, we can
always design a truncated water-filling scheme that obtains
optimal diversity by choosing $\beta \geq \beta_R$.

\begin{figure}
\begin{center}
\includegraphics[width  =1 \columnwidth]
        {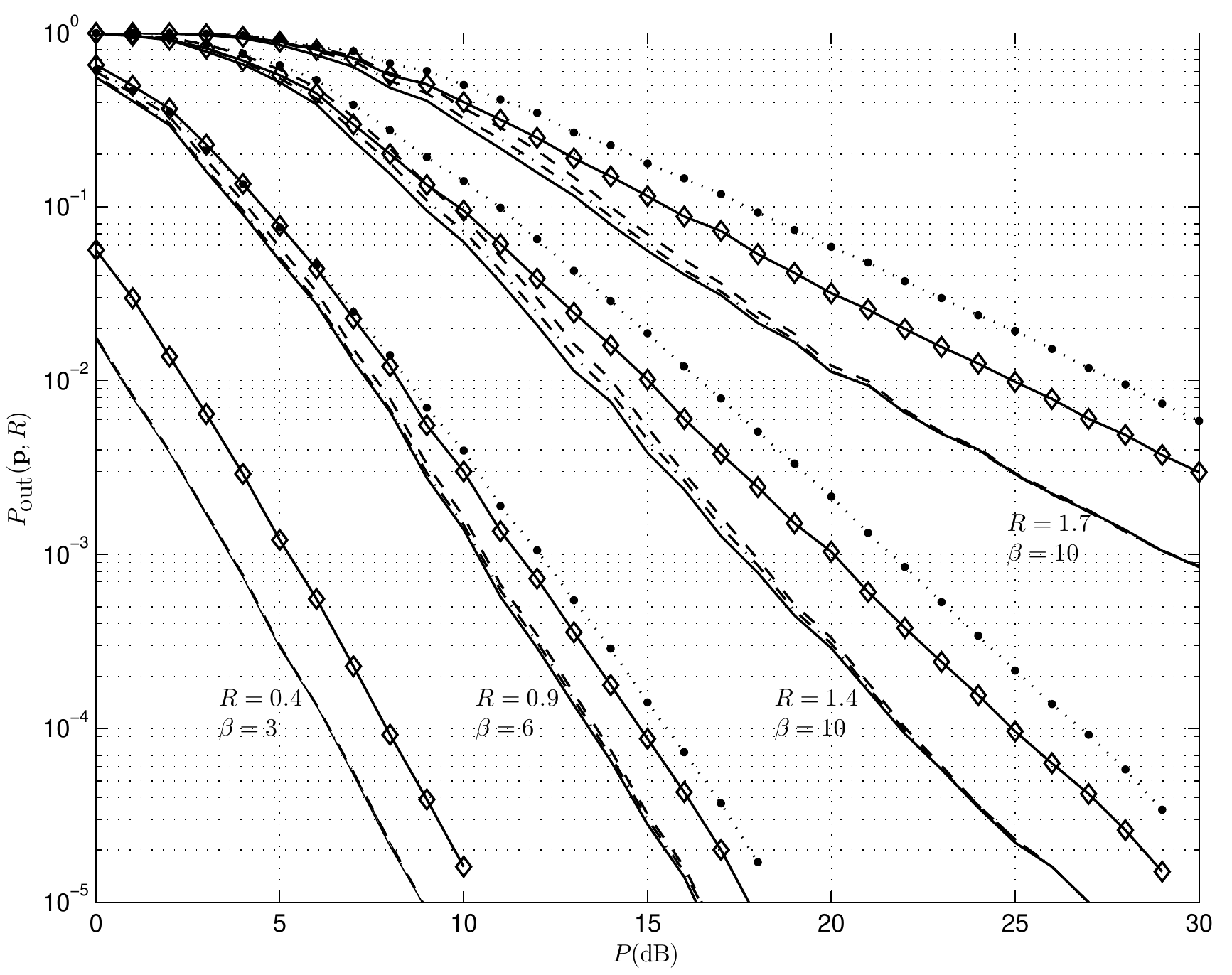}
\caption{Outage performance of various short-term power allocation schemes for
QPSK-input block-fading channels with $B=4$ and Rayleigh fading. The solid-lines
represent the optimal scheme; the solid lines with $\diamond$ represent uniform
power allocation; the dashed lines and dashed-dotted lines represent truncated
water-filling and its corresponding refinement, respectively; the dotted lines
represent the classical
water-filling scheme.} \label{fig:QPSK4blocksR0p5_0p9_1p4_1p7_beta10}
\end{center}
\end{figure}

With the results above, we choose $\beta$ as follows. For a
transmission rate $R$ that is not a discontinuity point of the
Singleton bound, we perform a simulation to compute the outage
probability obtained by truncated water-filling with various
$\beta \geq \beta_R$ and pick the $\beta$ that gives the best
outage performance. The dashed lines in Figure
\ref{fig:QPSK4blocksR0p5_0p9_1p4_1p7_beta10} illustrate the
performance of the obtained schemes for block-fading channels with
$B=4$ and QPSK input under Rayleigh fading. At all rates of
interest, the truncated water-filling schemes suffer only minor
losses in outage performance as compared to the optimal schemes
(solid lines), especially at high SNR. We also observe a
remarkable difference with respect to pure water-filling for
Gaussian inputs (dotted lines). As a matter of fact, pure
water-filling performs worse than uniform power allocation.

\begin{figure}
\begin{center}
\includegraphics[width =1\columnwidth]{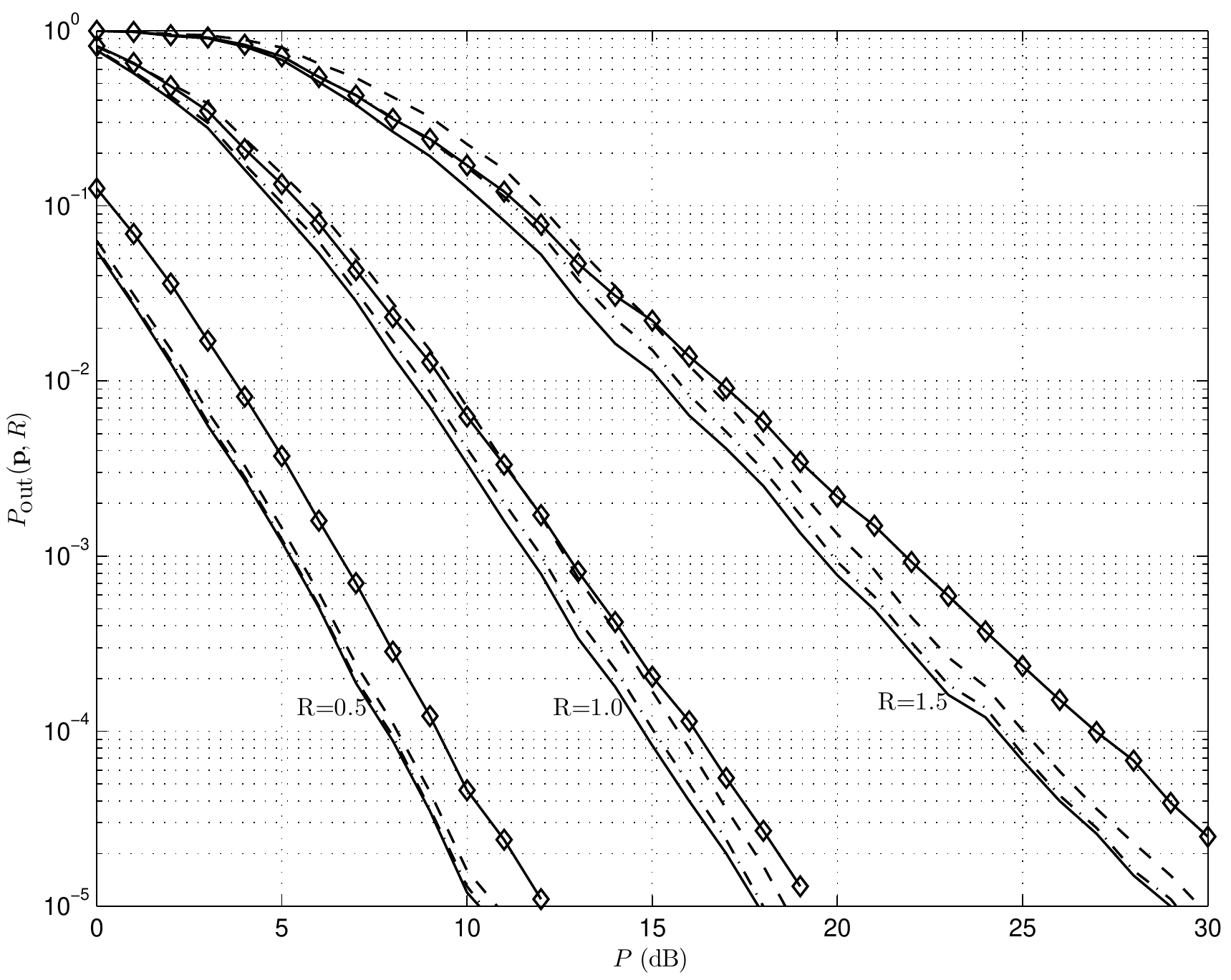}
\caption{Outage performance of various short-term power allocation schemes for
QPSK-input block-fading  channels with $B=4$ and Rayleigh fading. The solid-lines
  represent the optimal scheme; the solid lines with $\diamond$ represent the uniform
  power allocation; the dashed lines and dashed-dotted lines correspondingly
  represent the truncated water-filling and its refinement with $\beta = 15$.}
\label{fig:ref_trunc_wf_beta15}
\end{center}
\end{figure}

For rates at the discontinuous points of the Singleton bound, especially when
operating at high SNR, $\beta$ needs to be relatively large in order to
maintain diversity. However, large $\beta$ increases the gap between $I^{\rm
tw}(\rho)$ and $\IX(\rho)$, thus degrading the performance of the truncated
water-filling scheme. For $\beta=15$, the gap is illustrated by the dashed
lines in Figure \ref{fig:ref_trunc_wf_beta15}. In the extreme case where $\beta
\to \infty$, the truncated water-filling turns into the water-filling scheme,
which exhibits a significant loss in outage performance as illustrated in
Figure \ref{fig:QPSK4blocksR0p5_0p9_1p4_1p7_beta10}.

\subsubsection{Refined truncated water-filling schemes}
\label{se:refinement}
 To obtain a better approximation to the optimal power allocation scheme, we need
 a more accurate approximation to $\IX(\rho)$ in \eqref{eq:prob_max_cap}. We propose
 the following bound.

 \begin{equation}
 \label{eq:ref_target}
\IX(\rho)\leq I^{\rm ref}(\rho) \triangleq \begin{cases}
 \log_2(1+\rho), &\rho \leq \alpha\\
 \kappa \log_2(\rho)+ a, &\alpha < \rho \leq \beta \\
 \kappa \log_2(\beta)+ a, &\text{otherwise,}
 \end{cases}
 \end{equation}
where $\kappa$ and $a$ are chosen such that (in a dB scale)
$\kappa \log_2(\rho)+ a$ is a tangent to $\IX(\rho)$ at a
predetermined point $\rho_0$. Therefore $\alpha$ is chosen such
that $\kappa\log_2(\alpha)+a = \log_2(1+\alpha)$, and $\beta$ is a
design parameter. These parameters are reported in Table
\ref{tab:refined} for CM and BICM using various modulation
schemes. An example of the approximation is also illustrated by
the dashed-dotted curve in Figure \ref{fig:refine_bound_extended}.

The corresponding optimization problem can be written as
 \begin{equation}
\label{eq:ref_prob}
\begin{cases}
 \text{Maximize} &\sum_{b=1}^B  I^{\rm ref}(p_b \gamma_b)\\
 \text{Subject to} & \sum_{b=1}^B p_b \leq BP\\
 &p_b \geq 0, b=1, \ldots, B.
 \end{cases}
 \end{equation}
The refined truncated water-filling scheme $\mathbf{p}_{\rm
st}^{\rm ref}(\boldgamma)$ is given by the following theorem.
\begin{theorem}
\label{le:ref_sol} A solution to problem \eqref{eq:ref_prob} is
\begin{equation}
\pref_b = \frac{\beta}{\gamma_b}, \,\,\,b=1, \ldots, B
\end{equation}
if $\frac{1}{B}\sum_{b=1}^B \frac{\beta}{\gamma_b} < P$, and otherwise,
\begin{equation}
\pref_b = \begin{cases}
\frac{\beta}{\gamma_b}, &\eta \geq \frac{\beta}{\kappa \gamma_b}\\
\kappa \eta, &\frac{\alpha}{\kappa \gamma_b} \leq \eta <
\frac{\beta}{\kappa  \gamma_b}\\
\frac{\alpha}{\gamma_b}, &\frac{\alpha+1}{\gamma_b}\leq \eta <
\frac{\alpha}{\kappa\gamma_b}\\
\eta-\frac{1}{\gamma_b}, &\frac{1}{\gamma_b} \leq \eta <
\frac{\alpha+1}{\gamma_b}\\
0, &\text{otherwise,}
\end{cases}
\end{equation}
for $b = 1, \ldots, B$, where $\eta$ is chosen such that
\begin{equation}
\frac{1}{B}\sum_{b=1}^B \pref_b = P.
\end{equation}
\end{theorem}
\begin{proof}
See Appendix \ref{app:ref_sol}.
\end{proof}

The refined truncated water-filling scheme provides significant
gain over the truncated water-filling scheme, especially when the
transmission rate requires a relatively large $\beta$ to maintain
the outage diversity. The dashed-dotted lines in Figure
\ref{fig:ref_trunc_wf_beta15} show the outage performance of the
refined truncated water-filling scheme for block-fading channels
with $B=4$, and QPSK input under Rayleigh fading. The outage
performance of the refined truncated water-filling scheme is close
to the outage performance of the optimal case even at the rates
where the Singleton bound is discontinuous, i.e. rates $R= 0.5,
1.0, 1.5$. The performance gains of the refined scheme over the
truncated water-filling scheme at other rates are also illustrated
by the dashed-dotted lines in Figure
\ref{fig:QPSK4blocksR0p5_0p9_1p4_1p7_beta10}.

\section{Long-Term Power Allocation}
\label{se:long_term}

We consider systems with long-term power constraints, in
which the expectation of the power allocated to each block (over infinitely
many codewords) does not exceed $P$. This problem has been investigated in
\cite{CaireTariccoBiglieri1999} for block-fading channels with Gaussian
inputs. In this section, we obtain similar results for channels with discrete
inputs, and propose sub-optimal schemes that reduce the complexity of the
algorithm.

\subsection{Optimal Long-Term Power Allocation}
Following \cite{CaireTariccoBiglieri1999}, the problem can be formulated as
\begin{equation}
\label{eq:long_prob}
\begin{cases} \text{Minimize }& \Pr(I_B(\plt(\boldgamma),
\boldgamma)< R)\\
\text{Subject to}& \expectation{\mean{\plt(\boldgamma)}}{} \leq P,\\
\end{cases}
\end{equation}
where $\mean{\mathbf{p}} = \frac{1}{B}\sum_{b=1}^B p_b$.

The following theorem shows that the structure of the optimal long-term
solution $\pltopt(\boldgamma)$ of \cite{CaireTariccoBiglieri1999} for Gaussian
inputs is generalized to the discrete-input case.

\begin{theorem}
\label{theo:opt_longterm} Consider transmission over the
block-fading channel given in \eqref{eq:channel_model} with input
constellation $\mathcal{X}$. Assume that the power fading gains in
$\boldgamma$ follow the distribution given in
\eqref{eq:sq_Naka_dist}. Then, the optimal power allocation scheme
for systems with long-term constraint $P$,  is given by
\begin{equation}
\label{eq:long_term}
\pltopt(\boldgamma)= \begin{cases} \boldlp^{\rm opt} (\boldgamma),
&\mean{\boldlp^{\rm opt}(\boldgamma)} \leq s \\
0, &{\rm otherwise,}\end{cases}
\end{equation}
where
\begin{enumerate}
\item  $\boldlp^{\rm opt}(\boldgamma)$ is the solution of the
following optimization problem:
\begin{equation}
\label{eq:opt_min_pow_prob}
\left\{\begin{array}{ll}
{\rm Minimize} &\mean{\boldlp(\boldgamma)}\\
{\rm Subject \ to} & \frac{1}{B}\sum_{b=1}^B \IX(\lp_b \gamma_b) \geq R\\
& \lp_b \geq 0, b=1, \ldots, B.
\end{array}
\right.
\end{equation}
\item $s$ satisfies
\begin{equation}
\label{eq:def_s}
\left\{
\begin{array}{lll}
s &= \infty &$if$~ \lim_{s \to \infty} \expectation{\mean{\boldlp^{\rm
      opt}(\boldgamma)}}{\mathcal{R}(s)} \leq P \\
P&=\expectation{\mean{\boldlp^{\rm opt}(\boldgamma)}}{\boldgamma
\in \mathcal{R}(s)}, &{\rm
      otherwise,}
\end{array}\right.
\end{equation}
in which,
\begin{equation}
\label{eq:def_Rs}
\mathcal{R}(s) \triangleq \left\{\boldgamma \in \mathbb{R}_+^B:\mean{\boldlp^{\rm opt}
(\boldgamma)} \leq s\right\}.
\end{equation}
\end{enumerate}
\end{theorem}
\begin{proof}
See Appendix \ref{app:long_term_proof}.
\end{proof}

\begin{theorem}
\label{lem:opt_minpow_sol} Consider transmission at rate $R$ over
the block-fading channel given in \eqref{eq:channel_model} with
input constellation $\mathcal{X}$ and realized power fading gain
$\boldgamma$, the power allocation scheme that minimizes the input
power, which is a solution of the problem given in
\eqref{eq:opt_min_pow_prob}, is given by
\begin{equation}
\lp_b^{\rm opt}(\boldgamma) =
\frac{1}{\gamma_b}\MMSE^{-1}\left(\min\left\{1,
\frac{1}{\eta\gamma_b}\right\}\right), \label{eq:opt_minpow_sol}
\end{equation}
where $\eta$ is chosen such that
\begin{equation}
\frac{1}{B}\sum_{\substack{b=1\\ \gamma_b \geq \frac{1}{\eta}}}^B
\IX\left(\MMSE^{-1}\left(\frac{1}{\eta \gamma_b}\right)\right) =
R.
\end{equation}
\end{theorem}
\begin{proof}
See Appendix \ref{app:opt_minpow_sol}.
\end{proof}
As in the Gaussian input case \cite{CaireTariccoBiglieri1999},
the optimal power allocation scheme either transmits with the
minimum power that enables transmission at the target rate using
an underlying dual short-term scheme $\boldlp^{\rm
opt}(\boldgamma)$ with short-term constraint $\mean{\boldlp^{\rm
opt}(\boldgamma)} < s$, or turns off transmission (allocating zero
power) when the channel realization is bad. Therefore, there is no
power wastage on outage events.

\begin{figure}
\begin{center}
\includegraphics[width =1\columnwidth]{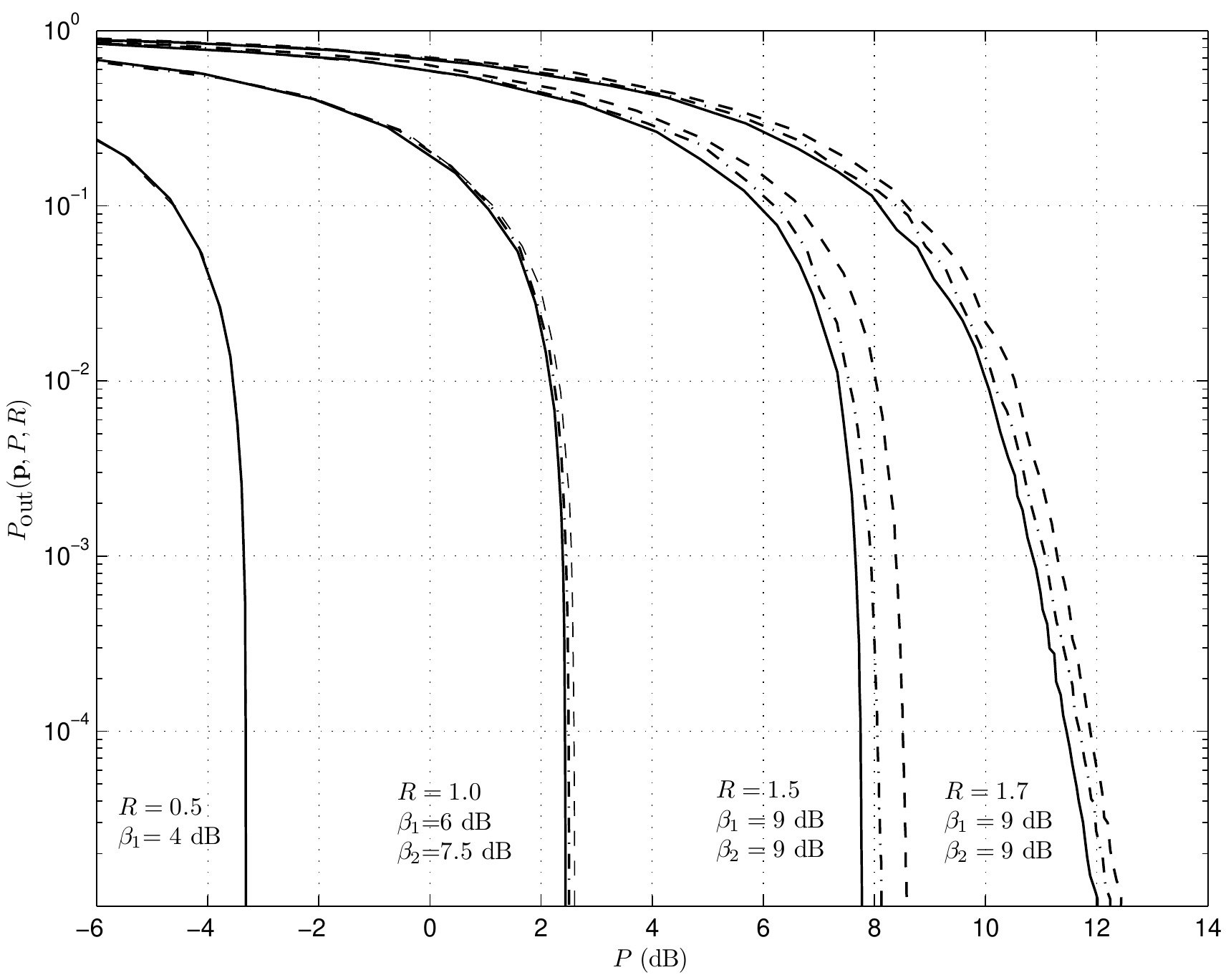}
\caption{Outage performance of various long-term power allocation schemes for
QPSK-input
  block-fading  channels with $B= 4$ and  Rayleigh fading. The solid-lines
  represent the optimal scheme; the dashed lines and dashed-dotted lines correspondingly
  represent the long-term truncated water-filling $\left(\mathbf{p}_{\rm
    lt}^{\rm tw}(\boldgamma) ~{\rm with}~ \beta_1 \right)$
  and its refinement $\left(\mathbf{p}_{\rm lt}^{\rm ref}(\boldgamma) {\ \rm
    with\ } \beta_2\right)$.}
\label{fig:long_schemes}
\end{center}
\end{figure}

\begin{figure}
\begin{center}
\includegraphics[width =1\columnwidth]{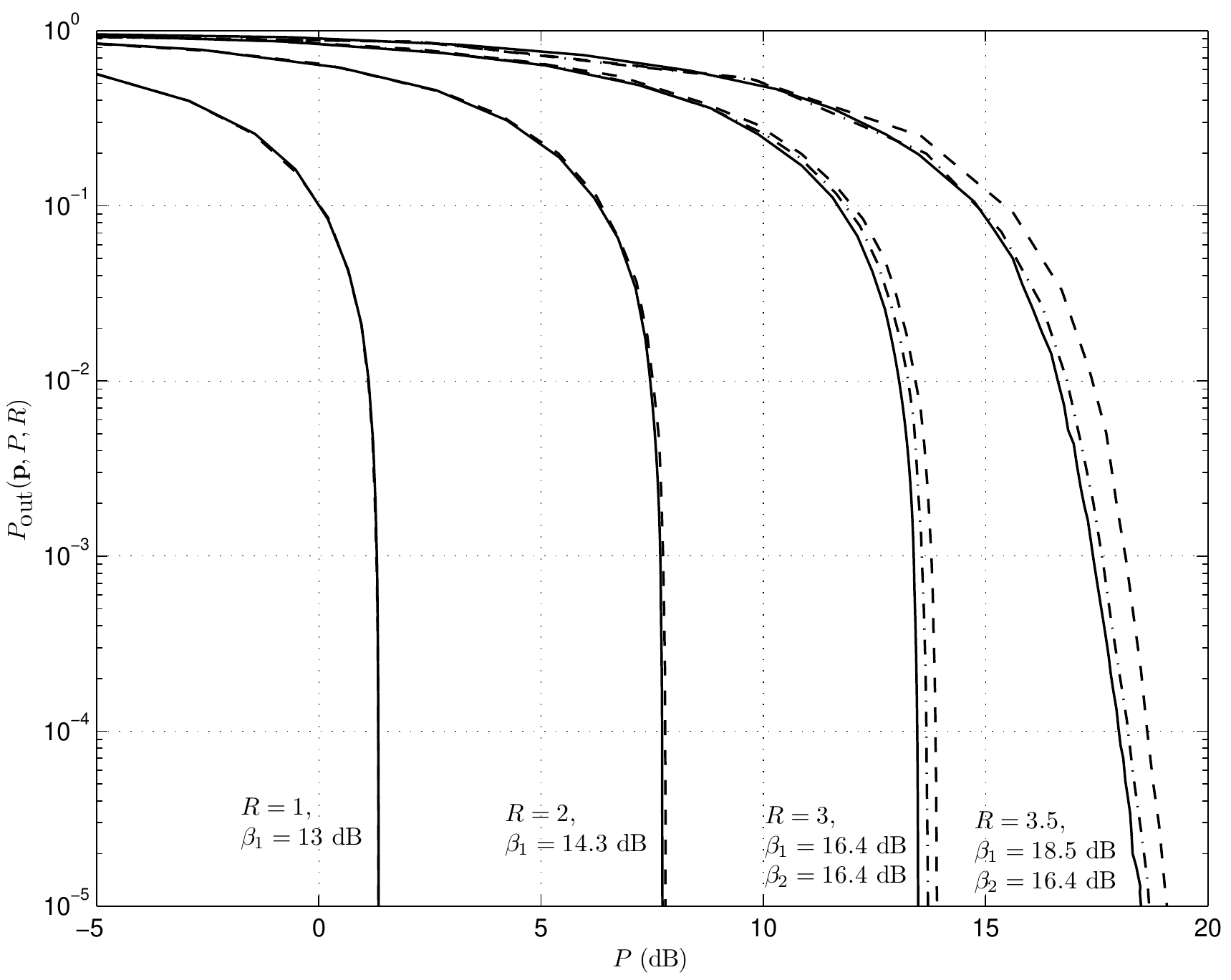}
\caption{Outage performance of various long-term power allocation schemes for
16-QAM-input
  4-block block-fading  channels under Rayleigh fading. The solid-lines
  represent the optimal scheme; the dashed lines and dashed-dotted lines correspondingly
  represent the long-term truncated water-filling $\left( \mathbf{p}_{\rm
    lt}^{\rm tw}(\boldgamma) ~{\rm with} ~ \beta_1 \right)$
  and its refinement $\left(\mathbf{p}_{\rm lt}^{\rm ref}(\boldgamma)~ {\rm with}~
  \beta_2 \right)$.}
\label{fig:long_schemes_QAM}
\end{center}
\end{figure}

The solid lines in Figures \ref{fig:long_schemes} and
\ref{fig:long_schemes_QAM} illustrate the outage performance of
optimal long-term power allocation schemes for transmission over
4-block block-fading channels with QPSK and $16$-QAM inputs and
Rayleigh fading ($m=1$). The simulation results suggest that for
transmission rates where $d_B(R)> 1$, zero outage probability can
be obtained with finite power. This agrees with the results
obtained for block-fading channels with Gaussian inputs
\cite{CaireTariccoBiglieri1999}, where only for $B>1$ zero outage
is possible.

To provide more insight into this effect, consider the following
long-term power allocation scheme,
\begin{equation}
\label{eq:app_long_contfad}
\plt(\boldgamma) = \begin{cases}
\boldlp(\boldgamma), &\boldgamma\in \mathcal{R}(s)\\
0, &{\rm otherwise,}
\end{cases}
\end{equation}
where $\boldlp(\boldgamma)$ is an arbitrary underlying short-term
power allocation scheme,
\begin{equation}
\mathcal{R}(s) = \left\{\boldgamma \in \mathbb{R}_+^B:
\mean{\boldlp(\boldgamma)} \leq s \right\},
\end{equation}
and $s$ is chosen to satisfy the long-term power constraint,
\begin{equation}
\label{eq:app_long_def_s} \expectation{\mean{\plt(\boldgamma)}}{} =
\expectation{\mean{\boldlp(\boldgamma)}}{\boldgamma \in \mathcal{R}(s)} = P.
\end{equation}
Assuming that for any $s$, $\boldlp(\boldgamma)$ satisfies
$\frac{1}{B}\sum_{b=1}^B \IX(\lp_b \gamma_b) \geq R$ for any
$\boldgamma\in\mathcal{R}(s)$, then the resulting outage probability of the
corresponding long-term power allocation is
\begin{equation}
 \Pout(\plt(\boldgamma), P(s),R) = \Pr(\boldgamma \notin \mathcal{R}(s)).
\end{equation}
For any long-term power constraint $P$, the long-term power allocation scheme
in \eqref{eq:app_long_contfad} depends on the threshold $s$ defined in
\eqref{eq:app_long_def_s}. Conversely, for any choice of the threshold
$s$, the long-term power $P(s)$ is given by
\begin{equation}
\label{eq:app_Ps}
P(s) = \expectation{\mean{\boldlp(\boldgamma)}}{\boldgamma \in \mathcal{R}(s)}.
\end{equation}
We now consider the behavior of the average power $P(s)$ and the
corresponding outage probability $\Pout(\plt(\boldgamma), P(s),R)$
when $s \to \infty$. In particular, we study the long-term
exponent defined as
\begin{equation}
d_{\rm lt}(R) \triangleq \lim_{P(s)\to \infty} -\frac{\log
\Pout(\plt(\boldgamma), P(s),R)}{\log P(s)}.
\end{equation}
Firstly, consider the following asymptotic relationship
between $s$ and $P(s)$.
\begin{proposition}
\label{prop:diff_Ps} Consider transmission over a block-fading channel with a
long-term power allocation scheme corresponding to an arbitrary underlying
short-term scheme $\boldlp(\boldgamma)$, a threshold $s$ given in
\eqref{eq:app_long_contfad}, and a long-term power constraint $P(s)$ given in
\eqref{eq:app_Ps}. Assume that $\boldlp(\boldgamma)$ is chosen such that
asymptotically in $s$, the outage probability satisfies
\begin{equation}
\label{eq:app_short_long_asym} \Pout(\plt(\boldgamma), P(s), R) \doteq
\mathcal{K} s^{-d(R)}
\end{equation}
for some finite $d(R) > 0$. Then,
\begin{equation}
\frac{d}{ds} P(s) \doteq \mathcal{K} d(R) s^{-d(R)}
\end{equation}
\end{proposition}
\begin{proof}
See Appendix \ref{se:long_asym}.
\end{proof}
From the previous proposition, we obtain the following result.
\begin{theorem}
\label{theo:long_term_asymp} Consider transmission over a block-fading channel
with a long-term power allocation scheme corresponding to an arbitrary
underlying short-term scheme $\boldlp(\boldgamma)$, a threshold $s$ given in
\eqref{eq:app_long_contfad}, and a long-term power constraint $P(s)$ given in
\eqref{eq:app_Ps}. Assume that $\boldlp(\boldgamma)$ is chosen such that
asymptotically in $s$, the outage probability satisfies
\begin{equation}
\Pout(\plt(\boldgamma), P(s), R) \doteq \mathcal{K} s^{-d(R)}
\end{equation}
for some finite $d(R) > 0$.
Then, if $d(R) > 1$ we have that
\[
\lim_{s \to \infty} P(s) = P_{\rm th}< \infty \,\,\,\, \text{and}
\,\,\,\, d_{\rm lt}(R) = \infty,
\]
while if $d(R) < 1$,
\begin{equation}
\label{eq:rel_short_long_diver} d_{\rm lt}(R) =
\frac{d(R)}{1-d(R)}.
\end{equation}
\end{theorem}
\begin{proof}
See Appendix \ref{se:long_asym}.
\end{proof}

The previous results highlight the effect of the power constraint on the outage
performance obtained by a specific power allocation scheme $\plt(\boldgamma)$.
In particular, $\Pout(\plt(\boldgamma), P(s), R)$ is the outage probability of
the block-fading channel with power allocation scheme $\boldlp(\boldgamma)$ and
short-term power constraint $\mean{\boldlp(\boldgamma)} < s$, and $d(R)$ is the
corresponding {\em short-term outage diversity}. When a long-term power
constraint is applied, the {\em long-term outage diversity} $d_{\rm lt}(R)$ is
affected in the following way:
\begin{itemize}
\item If $d(R)> 1$, then $d_{\rm lt}(R) = \infty$ and $\lim_{s \to \infty}
  P(s) = P_{\rm th} < \infty$. Therefore, since
\begin{equation}
\lim_{s \to \infty} \Pout(\plt(\boldgamma), P(s), R) = 0,
\end{equation}
there exists a threshold long-term power constraint $P_{\rm th}$
beyond which strictly zero outage probability is achieved, proving
that vanishing error probability can be achieved and that reliable
communication is possible at rates below $R$. $R$ is therefore a
lower bound to the {\em delay limited} capacity
\cite{HanlyTse1998} of the block-fading channel with power
constraint $P_{\rm th}$.

\item If $d(R)< 1$, then \eqref{eq:rel_short_long_diver} gives the relationship
between the {\em long-} and {\em short-term outage diversity}.
\end{itemize}


In order to apply the previous theorem to analyze the asymptotic
behaviour of the optimal power allocation for systems with
long-term constraints, we consider the following duality between
$\pst^{\rm opt}(\boldgamma)$ and $\boldlp^{\rm opt}(\boldgamma)$.
\begin{proposition}
\label{prop:dual_minpow_maxrate} Consider transmission at rate $R$ over the
block-fading channel given in \eqref{eq:channel_model} with the optimal
long-term power allocation scheme $\pltopt(\boldgamma)$ given in
\eqref{eq:long_term} and a long-term power constraint $P(s)$. Then,
independent of the fading statistics, the outage probability satisfies
\begin{equation}
\Pout(\plt^{\rm opt}(\boldgamma),P(s), R) = \Pout(\pst^{\rm opt}(\boldgamma),
s, R),
\end{equation}
where $\pst^{\rm opt}(\boldgamma)$ is the optimal power allocation scheme with
short-term power constraint $\mean{\pst^{\rm opt}(\boldgamma)} \leq s$.
\end{proposition}
\begin{proof}
See Appendix \ref{se:long_asym}.
\end{proof}

From Theorem \ref{theo:long_term_asymp}, we have the following result.

\begin{corollary}
Consider transmission at rate $R$ over the block-fading channel
given in \eqref{eq:channel_model} with the optimal long-term power
allocation scheme $\pltopt(\boldgamma)$. Assume input
constellation $\mathcal{X}$ of size $|\mathcal{X}|= 2^M$. Further
assume that the power fading gain $\boldgamma$ follows a
Nakagami-$m$ distribution given in \eqref{eq:sq_Naka_dist}. Then,
the delay-limited capacity is non-zero whenever $d_B(R)>
\frac{1}{m}$. Conversely, when $d_B(R)< \frac{1}{m}$, the outage
probability asymptotically behaves as
\begin{equation}
\Poutlt(\pltopt(\boldgamma), P(s), R) \doteq \mathcal{K}_{\rm
lt}^{\rm opt}  P^{-d_{\rm lt}^{\rm opt}(R)},
\end{equation}
where $P$ is the long-term power constraint, and $d_{\rm lt}^{\rm opt}(R)$ is
the optimal long-term outage diversity given by
\begin{equation}
d_{\rm lt}^{\rm opt}(R) = \frac{m d_B(R)}{1-md_B(R)}.
\end{equation}
\end{corollary}

\begin{proof}
From Propositions \ref{le:opt_diver} and \ref{prop:dual_minpow_maxrate}, we have
\begin{equation}
\Poutlt(\pltopt(\boldgamma), P(s), R) \doteq \mathcal{K}_{\rm opt} s^{-md_B(R)}.
\end{equation}
Therefore, the corollary can be obtained as a direct application of Theorem
\ref{theo:long_term_asymp}.
\end{proof}

\begin{figure}
\begin{center}
\includegraphics[width =1\columnwidth]{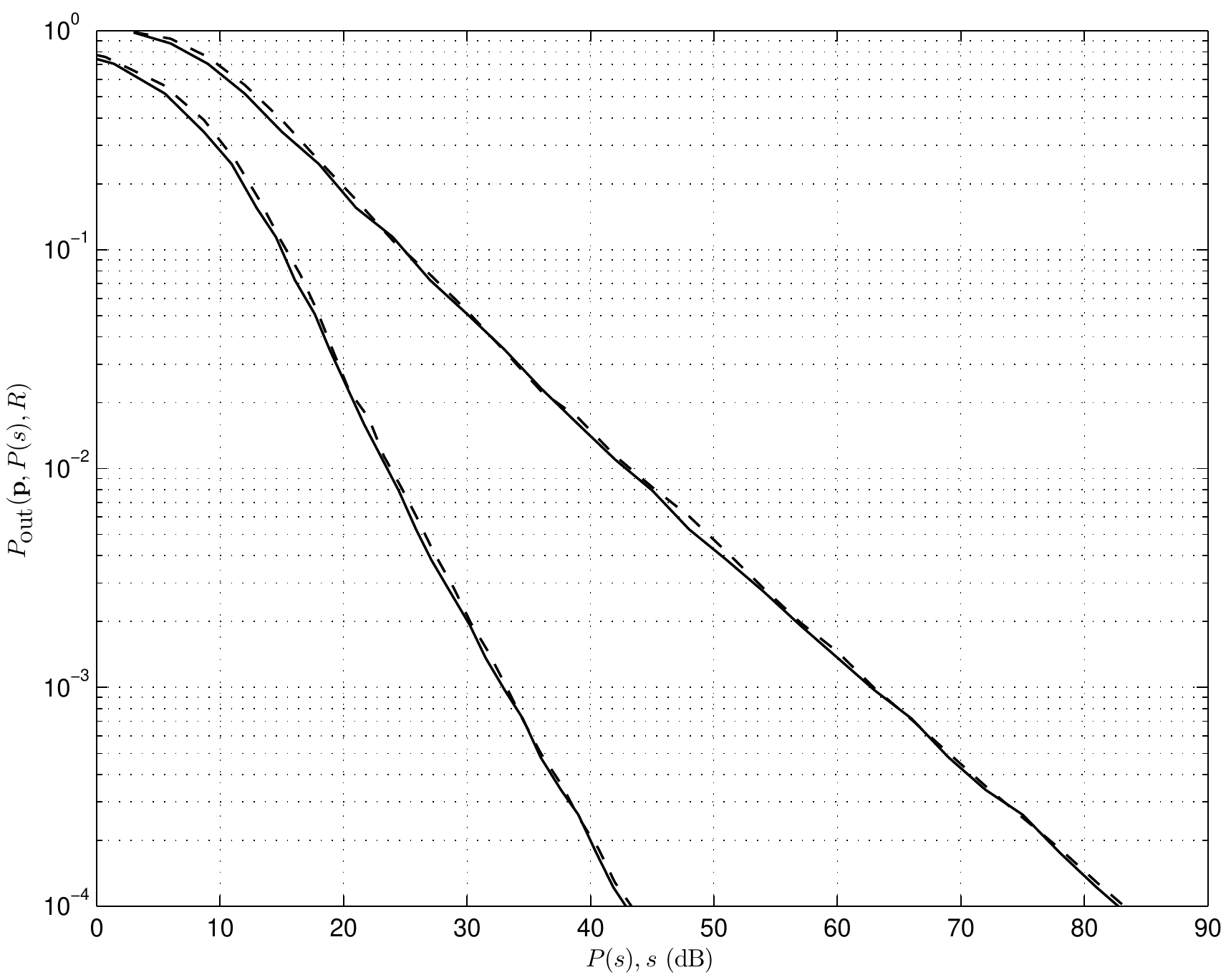}
\caption{Outage performance of long-term power allocation schemes
as a function of $P(s)$ and $s$ (dual short term scheme) with
QPSK-inputs in a block-fading channel with $B=4$, $m=0.5$, $R=1.7$ and
Rayleigh fading where $d_B(R)<\frac{1}{m}$. Solid-lines correspond
to the optimal scheme and dashed lines  correspond to truncated
water-filling with $\beta_1=3$. The curves as a function of $P(s)$
are the leftmost.} \label{fig:short_long_R1p7_m0p5}
\end{center}
\end{figure}

This behavior is illustrated in Figure
\ref{fig:short_long_R1p7_m0p5}, where the outage probability with
QPSK inputs, $m=0.5$ and $R=1.7$ has been plotted as a function of the
average long-term power $P(s)$ and as a function of the dual
short-term constraint $s$. As predicted by the previous results,
the dual short-term curve has slope $md_B(R) = 0.5$. Furthermore,
since $d_B(R)=1<\frac{1}{m}=2$, we observe that the long-term
outage curve (as a function of $P(s)$) has slope  $d_{\rm lt}^{\rm
opt}(R) = \frac{m d_B(R)}{1-md_B(R)}=1$.

\subsection{Sub-optimal Long-Term Power Allocation}

In the optimal long-term power allocation scheme $\plt^{\rm opt}(\boldgamma)$
given in Theorem \ref{theo:opt_longterm}, $s$ can be evaluated
offline for any fading distribution. Therefore, given an allocation scheme
$\boldlp^{\rm opt}(\boldgamma)$, the complexity required to evaluate $\plt^{\rm
opt}(\boldgamma)$ is low. Thus, the complexity of the long-term power
allocation scheme is mainly due to the complexity of evaluating $\boldlp^{\rm
  opt}(\boldgamma)$, which requires the evaluation or storage of $\MMSE(\rho)$ and
$\IX(\rho)$. In this section, we propose sub-optimal long-term power allocation
schemes by replacing the optimal underlying short-term algorithm $\boldlp^{\rm
opt}(\boldgamma)$ with simpler allocation rules.

A long-term power allocation scheme $\plt(\boldgamma)$
corresponding to an arbitrary $\boldlp(\boldgamma)$ is obtained by
replacing $\boldlp^{\rm opt}(\boldgamma)$ in \eqref{eq:long_term},
\eqref{eq:def_s} and \eqref{eq:def_Rs} with $\boldlp(\boldgamma)$.
From \eqref{eq:long_term},  the long-term power allocation scheme
$\plt(\boldgamma)$ satisfies
\begin{align}
\expectation{\mean{\plt(\boldgamma)}}{}=
&\expectation{\mean{\boldlp(\boldgamma)}}{\boldgamma \in
\mathcal{R}(s)} =P.
\end{align}
Therefore, a long-term power allocation scheme corresponding to an arbitrary
$\boldlp(\boldgamma)$ is sub-optimal with respect to $\plt^{\rm
  opt}(\boldgamma)$. Following the transmission strategy in the optimal scheme,
we consider the power allocation schemes $\boldlp(\boldgamma)$ that satisfy the
rate constraint $I_B(\boldlp(\boldgamma), \boldgamma)\geq R$ to avoid wasting
power on outage events. These schemes are sub-optimal solutions of problem
\eqref{eq:opt_min_pow_prob}. Based on the short-term schemes, two simple rules
are discussed in the next subsections.

\subsubsection{Long-term truncated water-filling scheme}
\label{se:tw_long}
 Similar to the short-term truncated water-filling scheme,
we consider approximating $\IX(\rho)$ in \eqref{eq:opt_min_pow_prob} by $I^{\rm
tw}(\rho)$ in \eqref{eq:target_tw}, which results in the following problem
\begin{equation}
\label{eq:tw_min_pow_prob}
\begin{cases}
\text{Minimize} &\mean{\boldlp(\gamma)}\\
\text{Subject \ to \ }&\frac{1}{B}\sum_{b=1}^B I^{\rm tw}(\lp_b \gamma_b)\geq R\\
&\lp_b \geq 0, b=1, \ldots, B
\end{cases}
\end{equation}

\begin{theorem}
\label{lem:tw_minpow_sol}
Problem \eqref{eq:tw_min_pow_prob} is solved by $\boldlp(\boldgamma)$ given as
\begin{equation}
\label{eq:ptw_org}
\lp_b = \min\left\{\frac{\beta}{\gamma_b}, \left(\eta -
  \frac{1}{\gamma_b}\right)_+ \right\}, b=1, \ldots, B,
\end{equation}
where $\eta$ is chosen such that
\begin{equation}
\frac{1}{B}\sum_{b=1}^B \log_2(1+\lp_b \gamma_b) = R.
\end{equation}
\end{theorem}
\begin{proof}
See Appendix \ref{app:tw_minpow_sol}.
\end{proof}
Note that since $I^{\rm tw}(\rho)$ is an upper bound on
$\IX(\rho)$, $\boldlp(\boldgamma)$ does not satisfy the rate
constraint $I_B(\boldlp(\boldgamma), \boldgamma)\geq R$. By
adjusting $\eta$, we can obtain a sub-optimal $\boldlp^{\rm
tw}(\boldgamma)$ as follows
\begin{equation}
\label{eq:lp_tw} \lp^{\rm tw}_b = \min\left\{\frac{\beta}{\gamma_b}, \left(\eta-
  \frac{1}{\gamma_b}\right)_+ \right\}, b=1, \ldots, B,
\end{equation}
where now $\eta$ is chosen according to the true mutual information with
discrete inputs, namely, we choose $\eta$ such that
\begin{equation}
\frac{1}{B}\sum_{b=1}^B \IX(\lp^{\rm tw}_b \gamma_b) = R.
\end{equation}
Using this scheme, we obtain a power allocation $\plt^{\rm
tw}(\boldgamma)$, which is the long-term power allocation scheme
corresponding to the sub-optimal $\boldlp^{\rm tw}(\boldgamma)$ of
$\boldlp^{\rm opt}(\boldgamma)$. The performance of the scheme is
illustrated by the dashed lines in Figures \ref{fig:long_schemes}
and \ref{fig:long_schemes_QAM}. As we observe,  the performance of
the truncated water-filling scheme is very close to that of the
optimal scheme.


Similar to the optimal power allocation scheme, we have the following duality
between $\pst^{\rm tw}(\boldgamma)$ and $\boldlp^{\rm
  tw}(\boldgamma)$.
\begin{proposition}
\label{prop:dual_tw_minpow_maxrate} Consider transmission at rate $R$ over the
block-fading channel given in \eqref{eq:channel_model} with power allocation
scheme $\plt^{\rm tw}(\boldgamma)$ and long-term power constraint $P(s)$. Then,
independent of the fading statistics, the outage probability satisfies
\begin{equation}
\Pout(\plt^{\rm tw}, P(s), R) = \Pout(\pst^{\rm tw}(\boldgamma), s, R),
\end{equation}
where $\pst^{\rm tw}(\boldgamma)$ is the truncated water-filling power
allocation scheme with short-term constraint $\mean{\pst^{\rm tw}(\boldgamma)}
\leq s$.
\end{proposition}
\begin{proof}
See Appendix \ref{app:dual_tw}
\end{proof}

Therefore, from Theorem \ref{theo:long_term_asymp} and Lemma
\ref{le:tw_diver}, we have the following result.
\begin{corollary}
Let $\plt^{\rm tw}(\boldgamma)$ be the long-term power allocation
scheme corresponding to $\boldlp^{\rm tw}(\boldgamma)$ given in
\eqref{eq:lp_tw}. Consider transmission at rate $R$ over the
block-fading channel given in \eqref{eq:channel_model} with the
long-term power allocation scheme $\plt^{\rm tw}(\boldgamma)$.
Assume that the power fading gain $\boldgamma$ follows a
Nakagami-$m$ distribution given in \eqref{eq:sq_Naka_dist}. Then,
the corresponding delay-limited capacity is non-zero if
$d_{\beta}(R)> \frac{1}{m}$, where
\begin{equation}
d_{\beta}(R) = 1 + \left\lfloor B\left(1- \frac{R}{\IX(\beta)}\right)\right\rfloor.
\end{equation}
\end{corollary}
\begin{proof}
From Propositions \ref{prop:dual_tw_minpow_maxrate} and \ref{le:tw_diver}, we have
\begin{equation}
\Pout(\plt^{\rm tw}(\boldgamma), P(s), R) \doteq \mathcal{K}_{\rm tw}
s^{-d_{\rm tw}(R)},
\end{equation}
where $d_{\rm tw}(R) \geq d_{\beta}(R)$. Therefore, the proof follows as a
result of Theorem \ref{theo:long_term_asymp}.
\end{proof}

In Figure \ref{fig:short_long_R1p7_m0p5} we also show in dashed
lines the  corresponding long-term truncated water-filling outage
curves, and we observe the same asymptotic behavior as for the
optimal scheme.

\subsubsection{Refinement of the long-term truncated water-filling scheme}
\label{se:ref_long} In order to improve the performance of the sub-optimal
scheme, we approximate $\IX(\rho)$ by $I^{\rm ref}(\rho)$ given in
\eqref{eq:ref_target}. Replacing $\IX(\rho)$ in \eqref{eq:opt_min_pow_prob} by
$I^{\rm ref}(\rho)$, we have the following problem
\begin{equation}
\label{eq:ref_min_pow_prob} \begin{cases}
\text{Minimize} &\mean{\boldlp(\gamma)}\\
\text{Subject \ to \ }&\frac{1}{B}\sum_{b=1}^B I^{\rm ref}(\lp_b \gamma_b)\geq R\\
&\lp_b \geq 0, b=1, \ldots, B.
\end{cases}
\end{equation}

\begin{theorem}
\label{theo:ref_minpow}
The problem given in \eqref{eq:ref_min_pow_prob} is solved by $\boldlp(\boldgamma)$ given as
\begin{equation}
\label{eq:ref_minpow_sol}
\lp_b = \begin{cases}
\frac{\beta}{\gamma_b}, &\eta \geq \frac{\beta}{\kappa \gamma_b}\\
\kappa \eta, &\frac{\alpha}{\kappa \gamma_b} < \eta < \frac{\beta}{\kappa
  \gamma_b}\\
\frac{\alpha}{\gamma_b}, &\frac{\alpha+1}{\gamma_b} \leq \eta \leq
\frac{\alpha}{\kappa \gamma_b}\\
\eta-\frac{1}{\gamma_b}, &\frac{1}{\gamma_b} \leq \eta <
\frac{\alpha+1}{\gamma_b}\\
0, &{\rm otherwise,}
\end{cases}
\end{equation}
where $\eta$ is chosen such that
\begin{equation}
\sum_{b=1}^B I^{\rm ref}(\lp_b \gamma_b) = BR.
\end{equation}
\end{theorem}
\begin{proof}
See Appendix \ref{app:ref_minpow_sol}.
\end{proof}
Following the arguments in Section \ref{se:tw_long}, we obtain the
sub-optimal $\boldlp^{\rm ref}(\boldgamma)$ of $\boldlp^{\rm
opt}(\boldgamma)$ from \eqref{eq:ref_minpow_sol} by choosing
$\eta$ in  such that
\begin{equation}
\frac{1}{B}\sum_{b=1}^B \IX(\lp^{\rm ref}_b \gamma_b) = R.
\end{equation}
%
%
The performance of the long-term power allocation corresponding to
$\boldlp^{\rm ref}(\boldgamma)$, $\plt^{\rm ref}(\boldgamma)$, is
illustrated by the dashed-dotted lines in Figures
\ref{fig:long_schemes} and \ref{fig:long_schemes_QAM}. We observe
that refined truncated water-filling leads to performance closer
to that of the optimal schemes than truncated water-filling. The
improvements are particularly clear for higher transmission rates.
\subsubsection{Approximation of $\IX(\rho)$}
The sub-optimal schemes in the previous sections are significantly
less complex than the corresponding optimal schemes, while only
suffering minor losses in outage performance. However, the
sub-optimal schemes still require the computation or storage of
$\IX(\rho)$ for determining $\eta$. This can be avoided by using
an approximation of $\IX(\rho)$. Let $\tilde{I}_{\mathcal
X}(\rho)$ be an approximation of $\IX(\rho)$ and $\Delta R$ be the
error measure given by
\begin{equation}
\label{eq:deltaR} \Delta R = \max_{\rho} \left\{\tilde{I}_{\mathcal X}(\rho)-
\IX(\rho)\right\}.
\end{equation}
Then, for a sub-optimal scheme $\boldlp(\boldgamma)$, $\eta$ is
chosen such that
\begin{equation}
\sum_{b=1}^B \tilde{I}_{\mathcal X}(\lp_b \gamma_b) = B(R+\Delta R)
\end{equation}
satisfies the rate constraint since
\begin{align}
\sum_{b=1}^B \IX(\lp_b \gamma_b) \geq \sum_{b=1}^B \tilde{I}_{\mathcal X}(\lp_b
\gamma_b)- B \Delta R = BR.
\end{align}

Following \cite{BrannstromRasmussenGrant2005}, we propose the following
approximation for $\IX(\rho)$
\begin{equation}
\tilde{I}_{\mathcal X}(\rho) = M\left(1-e^{-c_1 \rho^{c_2}}\right)^{c_3}.
\label{eq:approx_IX}
\end{equation}
For channels with QPSK input, using numerical optimization to
minimize the mean-squared-error between $\IX(\rho)$ and
$\tilde{I}_{\mathcal X}(\rho)$, we obtain the parameters, $c_1,
c_2, c_3$, shown in Table \ref{tab:approx_IX}. Using this
approximation to evaluate $\eta$ in subsections \ref{se:tw_long}
and \ref{se:ref_long}, we arrive at  computationally efficient
power allocation schemes with little loss in outage performance.

\begin{figure}
\begin{center}
\includegraphics[width =1\columnwidth]{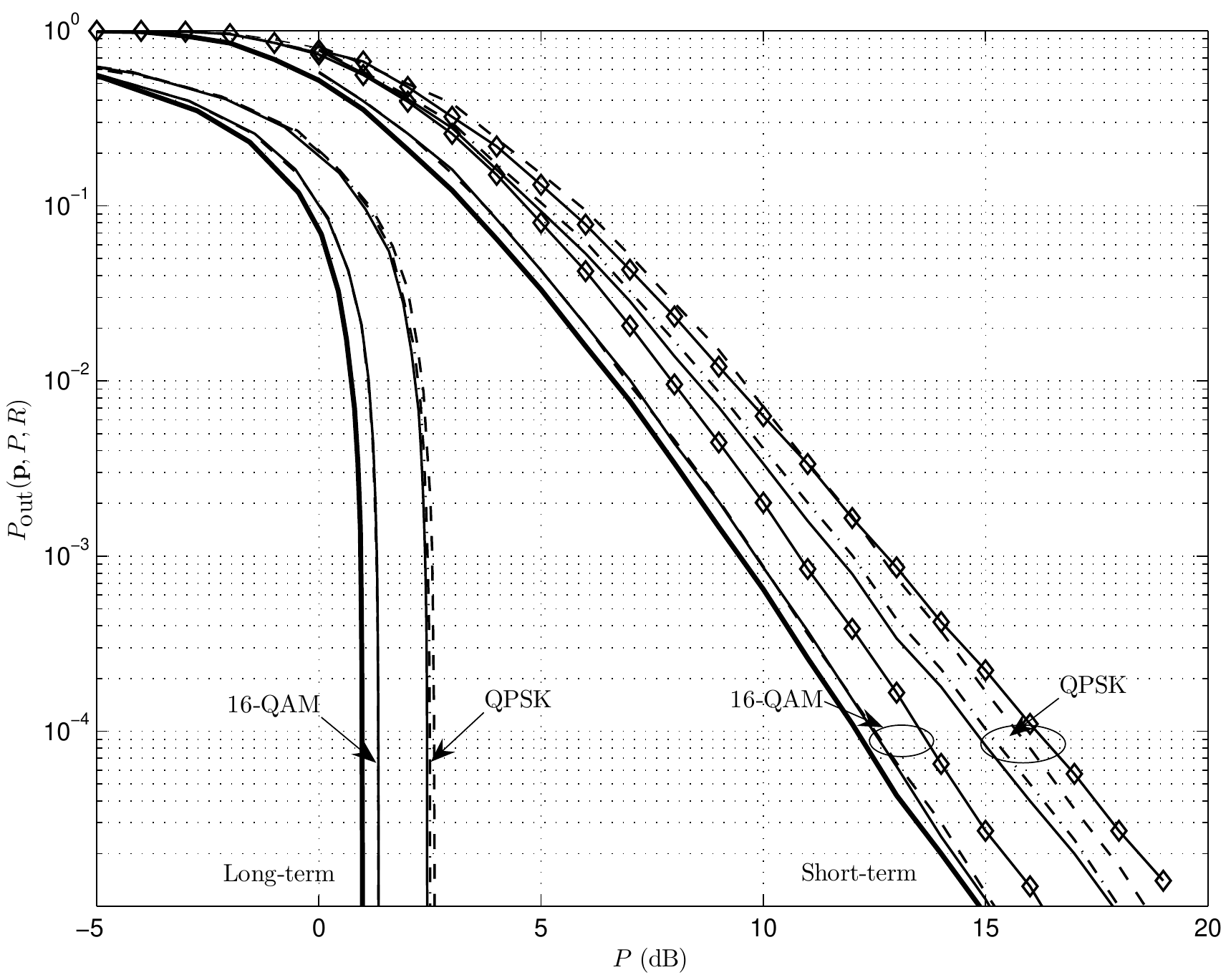}
\caption{Outage performance of various short- and long-term power
allocation schemes in a block-fading channel with $B=4$, $R=1$,
Rayleigh fading and Gaussian, QPSK and $16$-QAM inputs. The thick
solid line corresponds to the Gaussian input; the thin solid-lines
represent optimal scheme; the solid lines with $\diamond$
represent uniform power allocation; the dashed lines and
dashed-dotted lines represent truncated water-filling
($\mathbf{p}_{\rm lt}^{\rm tw}(\boldgamma)$ with $\beta_1=6$ dB
for QPSK input and 14.3 dB for QAM input respectively) and the
dashed dotted lines represents the corresponding refinement
($\mathbf{p}_{\rm lt}^{\rm ref}(\boldgamma)$ for QPSK input with
$\beta_2=5.5$).} \label{fig:comp_short_long}
\end{center}
\end{figure}

In Figure \ref{fig:comp_short_long}, we illustrate the significant
gains achievable by the long-term schemes when compared to
short-term schemes. As observed in
\cite{CaireTariccoBiglieri1999}, remarkable gains of $11$ dB at an
outage probability of $10^{-4}$ are possible with optimal
long-term power allocation when compared to uniform power
allocation  for Gaussian input distributions. As shown in Figure
\ref{fig:comp_short_long}, similar gains of the order of $12$ dB
at an outage probability of $10^{-4}$ are also achievable with
discrete inputs. Note that, due to the Singleton bound, the slope
of the QPSK-input short-term curves is not as steep as the slope
of the corresponding Gaussian input or $16$-QAM input curves. This
is due to the fact that both Gaussian and $16$-QAM inputs have SNR
exponent $d(R)=4$ while QPSK has $d_B(R)=3$. Figure \ref{fig:
comp_short_long_QAM_BICM} shows similar results comparing CM and
BICM.  In particular, the figure shows little loss between the
corresponding power allocation schemes. This is due to the fact
that the mutual information curves from CM and BICM with Gray
mapping do not differ much \cite{CaireTariccoBiglieri1998}. Once
again, in the case of BICM, the loss incurred by suboptimal
schemes is negligible.

\begin{figure}
\begin{center}
\includegraphics[width=1 \columnwidth]{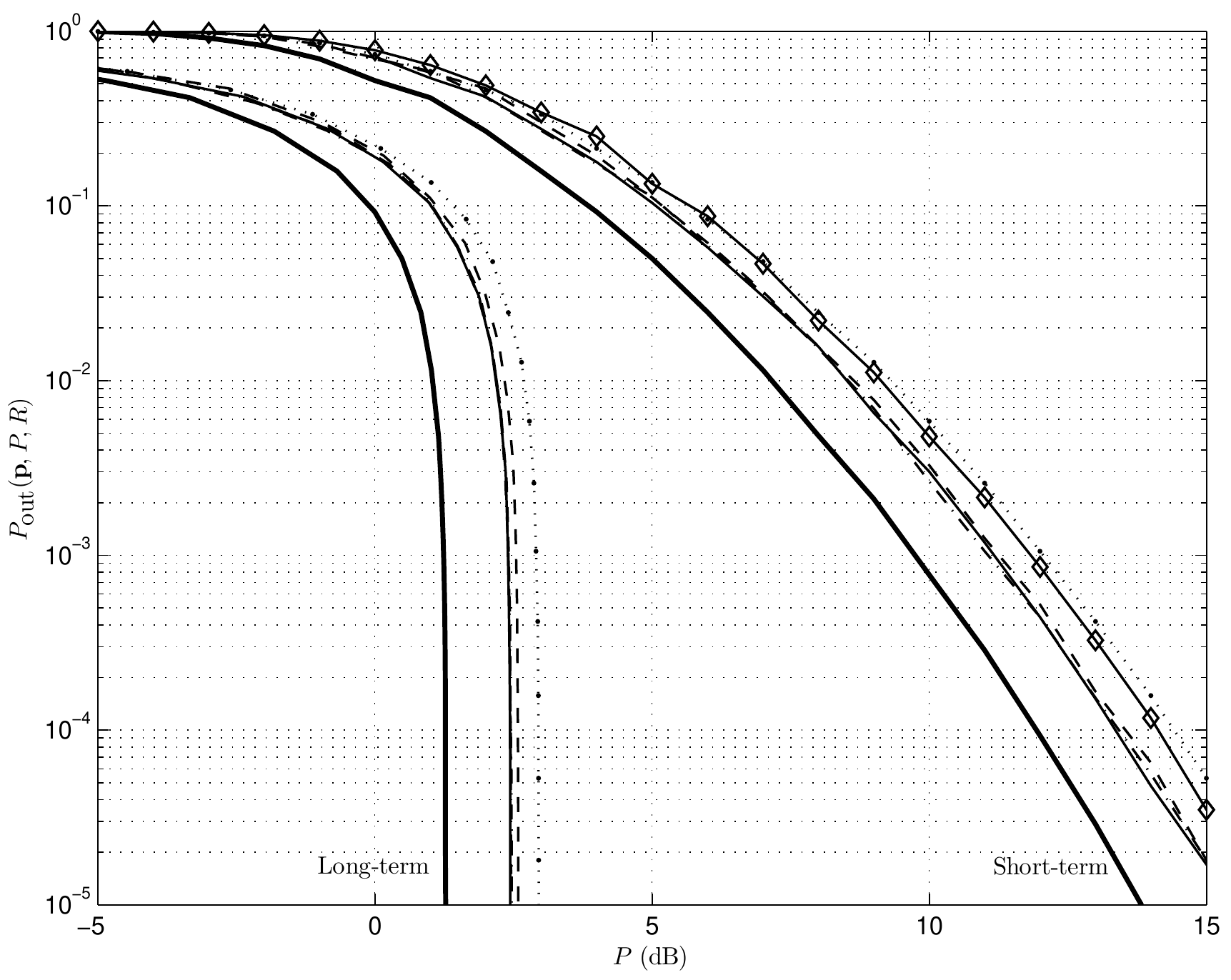}
\caption{Outage performance of various short- and long-term power
allocation schemes in an OFDM channels with $B= 64$ carriers, $R=
1$, Rayleigh fading and Gaussian, QPSK inputs. The thick solid
line corresponds to the Gaussian input, the thin solid lines
represent the optimal scheme; the solid lines  with $\diamond$
represent uniform power allocation; the dotted line represents the
pure water-filling; the dashed lines and   dashed-dotted lines
respectively represent long-term truncated water-filling
$\plt^{\rm tw}(\boldgamma)$and its corresponding refinement
$\plt^{\rm ref}(\boldgamma)$ with $\beta = 6$ dB. }
  \label{fig:ofdm_qpsk}
\end{center}
\end{figure}

We finally illustrate the application of the above results to
practical OFDM channels. In particular, we show in Figure
\ref{fig:ofdm_qpsk} the results corresponding to an OFDM channel
with $B=64$ sub-carriers, whose $9$-tap symbol-period-sampled
power delay profile is extracted from the ETSI BRAN-A model
\cite{etsi_brana} using a zero-hold order filter. The power delay
profile models a typical non-line-of-sight (NLOS) indoor office
scenario and is given in Table \ref{tab:brana}.  We observe a
similar behavior as in the block-fading channel.  In particular,
we show that in practical OFDM scenarios impressive gains of more
than $10$ dB with respect to uniform power allocation (eventually
reducing all outages) are possible. Note that due to the large
frequency diversity induced by the time-domain channel in Table
\ref{tab:brana}, the uniform and short-term power allocation
curves do not reveal their respective asymptotic slopes in the
error probability range shown in the figure.

\section{Conclusions}
\label{se:conclude}

We considered power allocation schemes for fixed-rate transmission
over discrete-input block-fading channels with transmitter and
receiver CSI under short- and long-term power constraints. We have
studied optimal and low-complexity sub-optimal schemes. In
particular, we have analyzed the optimal diversity orders and we
have shown that, in the long-term case, outages can be removed
provided that the short-term SNR exponent be greater than one. We have
illustrated the corresponding performances, showing significant
performance advantages on the order of $10$ dB of the proposed
long-term schemes when compared to uniform power allocation.
Furthermore, we have shown that minimal loss is incurred when
using the suggested sub-optimal schemes. We have also illustrated
the applicability and performance advantages of the proposed
techniques to practical OFDM situations.

\newpage
\appendices

\section{Optimal Power Allocation for Short-term Constraints}
\label{app:short_opt_proof}
\paragraph*{Proof of Lemma \ref{le:min_pout_sol}}
Since $\mathbf{p}^{\rm opt}_{\rm st}(\boldgamma)$ is the solution of
\eqref{eq:prob_max_cap}, we have
\begin{equation}
\sum_{b=1}^B \IX(\popt_b \gamma_b) \geq \sum_{b=1}^B \IX(p_b \gamma_b)
\end{equation}
for any power allocation scheme $\mathbf{p}_{\rm st}(\boldgamma)$
satisfying the short-term power constraint. Therefore,
\begin{equation}
\Pr\left(\sum_{b=1}^B \IX(\popt_b \gamma_b)< BR\right) \leq
\Pr\left(\sum_{b=1}^B \IX(p_b \gamma_b) < BR \right),
\end{equation}
and thus,
\begin{equation}
\Pout(\mathbf{p}^{\rm opt}_{\rm st}(\boldgamma), P,R)\leq \Pout(\mathbf{p}_b(\boldgamma),P, R)
\end{equation}
for any scheme $\mathbf{p}_{\rm st}(\boldgamma)$ satisfying the
short-term power constraint. This proves that $\mathbf{p}^{\rm
opt}_{\rm st}(\boldgamma)$ is a solution of
\eqref{eq:opt_short_prob}.
\endproof

~\\
\paragraph*{Proof of Proposition \ref{le:opt_diver}}
With the optimal power allocation scheme $\mathbf{p}^{\rm opt}_{\rm st}(\boldgamma)$,
the outage probability is given by
\begin{equation}
\Pout(\mathbf{p}^{\rm opt}_{\rm st}(\boldgamma),P, R) = \Pr\left(\frac{1}{B}\sum_{b=1}^B
I_{\mathcal{X}}(\popt_b\gamma_b)< R\right).
\end{equation}
Since $\mathbf{p}^{\rm opt}_{\rm st}(\boldgamma)$ is the solution of
\eqref{eq:prob_max_cap}, we have $\popt_b \geq 0, b=1, \ldots, B$ and
$\frac{1}{B}\sum_{b=1}^B \popt_b \leq P$. Therefore, $0 \leq \popt_b \leq BP, b=1,
\ldots, B$. Thus, $\Pout(\mathbf{p}^{\rm opt}_{\rm st}(\boldgamma),P, R)$ is lower bounded by
\begin{align}
\Pout (\mathbf{p}^{\rm opt}_{\rm st}(\boldgamma),P, R)& \geq
\Pr\left(\frac{1}{B}\sum_{b=1}^B I_{\mathcal{X}}(BP \gamma_b)< R\right)\\
&=\Pout (\peq(BP),PB, R),
\end{align}
namely, the outage probability of block-fading channels
corresponding to an equal allocation of power $PB$ per block.
Now, according to \cite{NguyenGuillenRasmussen2006}, under Nakagami-$m$
fading statistics, we have that
\begin{align}
\Pout(\mathbf{p}^{\rm opt}_{\rm st}(\boldgamma),P, R) &~\dot{\geq}~ \mathcal{K}
(BP)^{-md_B(R)}\\
&= \mathcal{K}B^{-md_B(R)}
P^{-md_B(R)}.
\label{eq:asym_lbound_opt}
\end{align}
Conversely, since the power allocation scheme is optimal, the
outage performance is upper bounded by the allocation scheme that
assigns power $P$ to each block. Therefore,
\begin{align}
\Pout(\mathbf{p}^{\rm opt}_{\rm st}(\boldgamma),P, R) &\leq \Pout(\peq(P),P, R) \\
&\doteq \mathcal{K}P^{-md_B(R)}.
\label{eq:asym_ubound_opt}
\end{align}
From \eqref{eq:asym_lbound_opt} and \eqref{eq:asym_ubound_opt}, we have
\begin{equation}
\Pout(\mathbf{p}^{\rm opt}_{\rm st}(\boldgamma),P, R) \doteq  \mathcal{K}_{\rm opt} P^{-m d_B(R)}.
\end{equation}
Thus, the diversity obtained by the optimal power allocation
scheme is given by $m d_B(R)$, which is the same as that of the
uniform power allocation scheme. This concludes the proof of the
Proposition.
\endproof
\newpage

\section{Proof of Theorem \ref{le:trunc_wfill_sol}}
\begin{proof}
\label{app:tw_sol} The power allocation algorithm of interest is the solution
of the optimization problem \eqref{eq:trunc_wfill_prob}. Since $f(p_b
\gamma_b)$ is constant at $\log_2(1+\beta)$ for $p_b \geq
\frac{\beta}{\gamma_b}$, having $p_b > \frac{\beta}{\gamma_b}$ does not
provides any gain to the target function in \eqref{eq:trunc_wfill_prob}.
Therefore, the solution of the following optimization problem
\begin{equation}
\label{eq:sim_trunc_wfill_prob}
\left\{\begin{array}{lll}
{\rm Minimize} &f_0(\mathbf{p}) \triangleq - \frac{1}{B}\sum_{b=1}^B \log_2(1+ p_b \gamma_b)\\
{\rm Subject \ to} & f_b(\mathbf{p}) \triangleq -p_b \leq 0,  &b = 1, \ldots, B\\
& g_b(\mathbf{p}) \triangleq p_b - \frac{\beta}{\gamma_b} \leq 0, &b=1, \ldots, B\\
&h(\mathbf{p}) \triangleq \sum_{b=1}^B p_b \leq BP
\end{array}\right.
\end{equation}
is also a solution of \eqref{eq:trunc_wfill_prob}.

It can be verified that the functions $f_0(\mathbf{p}),
f_b(\mathbf{p}), g_b(\mathbf{p}), b=1, \ldots, B$, and
$h(\mathbf{p})$ are convex. Therefore, according to the
Karush-Kuhn-Tucker (KKT) conditions  \cite{BoydVandenberghe2004},
the solution $\mathbf{p}^{\rm tw}$of
\eqref{eq:sim_trunc_wfill_prob} must satisfies
\begin{align}
\label{eq:discrete_cond1} \nu &\geq 0\\
\label{eq:discrete_cond2} \nu \left(\sum_{b=1}^B \ptwfill_b - BP\right) &=0\\
\label{eq:discrete_cond3}\sum_{b=1}^B \ptwfill_b - BP &\leq 0 \\
\label{eq:discrete_cond4}\ptwfill_b &\geq 0, &b=1, \ldots, B \\
\label{eq:discrete_cond5}\lambda_b &\geq 0, &b = 1, \ldots, B \\
\label{eq:discrete_cond6}-\lambda_b \ptwfill_b &=0, &b=1, \ldots, B\\
\label{eq:discrete_cond7}\alpha_b &\geq 0, &b = 1, \ldots, B \\
\label{eq:discrete_cond8}\ptwfill_b-\frac{\beta}{\gamma_b} &\leq 0, &b = 1, \ldots, B \\
\label{eq:discrete_cond9}\alpha_b \left(\ptwfill_b-\frac{\beta}{\gamma_b}\right) &=0, &b=1, \ldots, B\\
\label{eq:discrete_cond10}-\frac{ \gamma_b\log_2 e}{1+\ptwfill_b \gamma_b}-
\lambda_b +\alpha_b + \nu &=0, &b=1, \ldots, B,
\end{align}
where $\nu, \lambda_b, \alpha_b, b=1, \ldots, B$ are the Lagrangian multipliers.
For any $b$,
\begin{itemize}
\item If $\lambda_b >0$, from \eqref{eq:discrete_cond6},
$\ptwfill_b = 0$. Therefore, from \eqref{eq:discrete_cond9}
$\alpha_b = 0$. In this case condition \eqref{eq:discrete_cond10}
is satisfied only if $\nu > \gamma_b \log_2 e$. \item If
$\lambda_b = 0$, we have the following cases
\begin{itemize}
\item If $\alpha_b > 0$, from \eqref{eq:discrete_cond9},
$\ptwfill_b = \frac{\beta}{\gamma_b}$. In this case,
\eqref{eq:discrete_cond10} is satisfied only when $\nu <
\frac{\gamma_b \log_2 e}{\beta+1}$. \item If $\alpha_b= 0$, from
\eqref{eq:discrete_cond10}, $\ptwfill_b = \frac{\log_2 e}{\nu}-
\frac{1}{\gamma_b}$. Furthermore, from \eqref{eq:discrete_cond4}
and \eqref{eq:discrete_cond8}, we have $\frac{\gamma_b \log_2
e}{\beta+1} \leq \nu \leq \gamma_b \log_2 e$.
\end{itemize}
\end{itemize}
Therefore, for any choice of $\nu$, we must have
\begin{align}
 \ptwfill_b &= \left\{\begin{array}{ll}
\frac{\beta}{\gamma_b}, & \nu < \frac{\gamma_b \log_2 e}{\beta+1}\\
\frac{\log_2 e}{\nu} - \frac{1}{\gamma_b}, &\frac{\gamma_b \log_2 e }{\beta+1} \leq \nu \leq \gamma_b \log_2 e \\
0, &\rm{otherwise}
\end{array}\right.\\
\label{eq:discrete_pstar} &= \min\left\{\frac{\beta}{\gamma_b},
\left(\frac{\log_2 e}{\nu}- \frac{1}{\gamma_b}\right)_+\right\}.
\end{align}
The solution in \eqref{eq:discrete_pstar} satisfies conditions
\eqref{eq:discrete_cond4}--\eqref{eq:discrete_cond10}. We are left to choose
$\nu \geq 0$ such that conditions \eqref{eq:discrete_cond1} --
\eqref{eq:discrete_cond3} are satisfied. If $\nu = 0$, from
\eqref{eq:discrete_pstar}, $\ptwfill_b = \frac{\beta}{\gamma_b}, ~b=1, \ldots,
B$. Therefore, $\nu = 0$ is valid only if
\begin{equation}
\sum_{b=1}^B \frac{\beta}{\gamma_b} \leq BP.
\end{equation}
Otherwise, if $\sum_{b=1}^B \frac{\beta}{\gamma_b} > BP$, choose $\nu$
such that
\begin{equation}
\sum_{b=1}^B \ptwfill_b = BP.
\end{equation}

Therefore, letting $\eta = \frac{\log_2 e}{\nu}$, the  solution of
\eqref{eq:sim_trunc_wfill_prob} can be summarized as follows. If
$\sum_{b=1}^B \frac{\beta}{\gamma_b} \leq BP$,
\begin{equation}
\ptwfill_b = \frac{\beta}{\gamma_b}, b=1, \ldots, B.
\end{equation}
Otherwise, the solution is
\begin{equation}
\ptwfill_b = \min\left\{\frac{\beta}{\gamma_b}, \left(\eta-
\frac{1}{\gamma_b}\right)_+\right\},
\end{equation}
where $\eta$ is the solution of
\begin{equation}
\label{eq:water_level_discrete} \sum_{b=1}^B \min\left\{\frac{\beta}{\gamma_b},
\left(\eta- \frac{1}{\gamma_b}\right)_+\right\} = BP.
\end{equation}
\end{proof}

\newpage

\section{Proof of Proposition \ref{le:tw_diver}}
\label{app:proof_tw_dive}

\begin{proof}
With the truncated water-filling power allocation scheme $\pst^{\rm
  tw}(\boldgamma)$, the outage probability is given by
\begin{equation}
\Pout(\mathbf{\pst^{\rm tw}}(\boldgamma), P, R) = \Pr\left(\frac{1}{B}
\sum_{b=1}^B I_{\mathcal{X}}(\ptw_b \gamma_b)< R\right).
\end{equation}
The outage probability can be upper bounded by
\begin{equation}
\label{eq:upbound_tw_outage}
\Pout(\pst^{\rm tw}(\boldgamma), P, R) \leq
\Pr\left(\frac{1}{B}\sum_{b=1}^B I_{\mathcal X}^{\beta}(\ptw_b \gamma_b) <
R\right),
\end{equation}
where $I_{\mathcal X}^{\beta}(\rho)$ is a lower bound to $I_{\mathcal
  X}(\rho)$ given by
\begin{equation}
\label{eq:ind_mu_beta}
I_{\mathcal X}^{\beta}(\rho) = \left\{\begin{array}{ll}
I_{\mathcal X}(\beta), & \rho \geq \beta\\
0, &{\rm otherwise.}
\end{array}\right.
\end{equation}
We now further upper bound $\Pout (\pst^{\rm tw}(\boldgamma), P,
R)$ using the following proposition.
\begin{proposition}
\label{prop:bound_sum_mutual}
Consider the truncated water-filling scheme given in
\eqref{eq:trunc_wfill_sol}. For any channel realization $\boldgamma$, we have
\begin{equation}
\label{eq:bound_sum_mutual}
\sum_{b=1}^B I_{\mathcal X}^{\beta}(\ptw_b \gamma_b) \geq  \sum_{b=1}^B I_{\mathcal
  X}^{\beta}(P\gamma_b),
\end{equation}
where $I_{\mathcal X}^{\beta}(\rho)$ is given in \eqref{eq:ind_mu_beta}.
\end{proposition}
\begin{proof}
According to \eqref{eq:ind_mu_beta}, $I_{\mathcal X}^{\beta}(P \gamma_b)$ is
non-zero only if $\gamma_b \geq \frac{\beta}{P}$. Therefore, we need to prove
that if $\gamma_b \geq \frac{\beta}{P}$ then $\ptw_b \geq
\frac{\beta}{\gamma_b}$ for all realization of $\boldgamma$.
Without loss of
generality, assume that $\gamma_1 \leq \ldots \leq \gamma_B$. If $\gamma_B <
\frac{\beta}{P}$, \eqref{eq:bound_sum_mutual} is certainly true. Otherwise,
there exists a $k, 1 \leq k \leq B$, such that  $\gamma_{k-1} < \frac{\beta}{P} \leq \gamma_k
\leq \ldots \leq \gamma_B$. Consider the following two cases:
\begin{itemize}
\item If $\sum_{b=1}^B \frac{1}{\gamma_b}< \frac{BP}{\beta}$ then from
  \eqref{eq:trunc_wfill_sol}, $p_b^{\rm tw} = \frac{\beta}{\gamma_b}, ~b=1, \ldots, B$.
\item Otherwise, from \eqref{eq:trunc_wfill_sol}, the power allocation
  solution is given by
\begin{equation}
\label{eq:tw_sol_sim} \ptw_b = \min\left\{\frac{\beta}{\gamma_b},
\left(\eta-\frac{1}{\gamma_b}\right)_+\right\}, ~b=1, \ldots, B,
\end{equation}
where $\eta$ is chosen such that
\begin{equation}
\label{eq:tw_constraint}
\sum_{b=1}^B \ptw_b = BP.
\end{equation}
Since $\gamma_k \geq \frac{\beta}{P}$, we have  from \eqref{eq:tw_sol_sim}
\begin{equation}
\ptw_b \leq \frac{\beta}{\gamma_b} \leq \frac{\beta}{\gamma_k} \leq P, ~b=k,
\ldots, B.
\end{equation}
Therefore from \eqref{eq:tw_constraint},
\begin{equation}
\label{eq:contradict1}
\sum_{b=1}^k \ptw_b = \sum_{b=1}^B \ptw_b - \sum_{b=k+1}^B \ptw_b \geq kP.
\end{equation}
Now, suppose
\begin{equation}
\label{eq:assumption}
\eta < \frac{\beta+1}{\gamma_k},
\end{equation}
then, for $b=1, \ldots, k$,
\begin{equation}
\ptw_b \leq \eta-\frac{1}{\gamma_b} < \frac{\beta+1}{\gamma_k}-
\frac{1}{\gamma_k} = \frac{\beta}{\gamma_k} \leq P.
\end{equation}
Thus, $\sum_{b=1}^k \ptw_b < kP$, which contradicts to \eqref{eq:contradict1}.
Therefore, assumption \eqref{eq:assumption} is invalid. We then conclude that
$\eta \geq \frac{\beta+1}{\gamma_k} \geq \frac{\beta+1}{\gamma_b}, b=k, \ldots,
B$. Therefore from \eqref{eq:tw_sol_sim}, $\ptw_b = \frac{\beta}{\gamma_b}, b=k,
\ldots, B$.
\end{itemize}
Therefore, in all cases, we have $\ptw_b = \frac{\beta}{\gamma_b}$ if
$\gamma_b \geq \frac{\beta}{P}$. This concludes the proof of the proposition.
\end{proof}
From Proposition \ref{prop:bound_sum_mutual}, we can further upper bound
$\Pout(\pst^{\rm tw}(\boldgamma),P,  R)$ by
\begin{equation}
\label{eq:upbound_tw_outprob} \Pout(\pst^{\rm tw}(\boldgamma), P, R) \leq
\Pout^{\beta}(\mathbf{p}_{\rm eq}(P), P, R) \triangleq \Pr\left(\frac{1}{B}
\sum_{b=1}^B I_{\mathcal X}^{\beta}(P \gamma_b)< R \right).
\end{equation}

The asymptotic behavior of $\Pout^{\beta}(\mathbf{p}_{\rm eq}(P), P, R)$ is
given by the following proposition
\begin{proposition}
\label{prop:asym_poutind} Assume that $\gamma_b$ follows the
distribution given in \eqref{eq:sq_Naka_dist}, then
$\Pout^{\beta}(\mathbf{p}_{\rm eq}(P), P, R)$ in
\eqref{eq:upbound_tw_outprob} asymptotically behaves as
\begin{equation}
\Pout^{\beta}(\mathbf{p}_{\rm eq}(P), P, R) \doteq
\mathcal{K}_{\beta}P^{-md_{\beta}},
\end{equation}
where
\begin{equation}
d_{\beta} = 1+ \left\lfloor B\left(1-\frac{R}{I_{\mathcal
X}(\beta)}\right)\right\rfloor,
\end{equation}
and $I_{\mathcal X}(\rho)$ is the input-output mutual information of a AWGN
channel with input constellation $\mathcal{X}$ and SNR $\rho$.
\end{proposition}
\begin{proof}
Consider the random set given by
$\mathcal{S}_{\beta}=\left\{i \in \{1, \ldots, B\}: \gamma_i >
\frac{\beta}{P}\right\}$. Then for $b= 1, \ldots, B$,
\begin{equation}
\Pr(b \in \mathcal{S}_{\beta}) = \Pr\left(\gamma_b >
\frac{\beta}{P}\right) = 1-
F_{\gamma}\left(\frac{\beta}{P}\right) \triangleq p_{\beta}.
\end{equation}
The asymptotic behavior of $p_{\beta}$ is given by
\begin{align}
p_{\beta} &= \frac{\Gamma\left(m, m \frac{\beta}{\gamma_b}\right)}{\Gamma(m)}\\
&\doteq \frac{\Gamma(m)- \frac{1}{m}\left(m \frac{\beta}{P}\right)^m}
{\Gamma(m)}\\
1- p_{\beta} &\doteq  \frac{m^{m-1}\beta^m}{\Gamma(m)}P^{-m}.
\end{align}
Since $\gamma_1, \ldots, \gamma_B$ are independent random variables,
$|\mathcal{S}_{\beta}|$ is binomially distributed
\begin{align}
\Pr(|\mathcal{S}_{\beta}|= t)&= \binom{B}{t}p_{\beta}^t (1-p_{\beta})^{B-t}\\
&\doteq \binom{B}{t} \left(\frac{m^{m-1}\beta^m}{\Gamma(m)}\right)^{B-t}P^{-m(B-t)}.
\end{align}
Now from \eqref{eq:ind_mu_beta},
\begin{equation}
I_{\mathcal X}^{\beta} (P\gamma_b) = \left\{\begin{array}{ll}
I_{\mathcal X}(\beta), &b \in \mathcal{S}_{\beta}\\
0, &{\rm otherwise.}
\end{array}\right.
\end{equation}
Therefore,
\begin{align}
\Pout(P, R) &= \Pr\left(\sum_{b=1}^B I_{\mathcal X}^{\beta}(P \gamma_b)<
BR\right)\\
&= \Pr(|\mathcal{S}_{\beta}|I_{\mathcal X}(\beta)< BR)\\
&= \Pr\left(|\mathcal{S}_{\beta}|< \frac{BR}{I_{\mathcal X}(\beta)}\right)\\
&=\sum_{t=0}^{\left\lceil \frac{BR}{I_{\mathcal X}(\beta)}\right\rceil-1}
\Pr(|\mathcal{S}_{\beta}|= t)\\
\label{eq:sem_asym_poutind}
&\doteq \sum_{t=0}^{\left\lceil\frac{BR}{I_{\mathcal X}(\beta)}\right\rceil-1}
\binom{B}{t} \left(\frac{m^{m-1}\beta^m}{\Gamma(m)}\right)^{B-t}P^{-m(B-t)}.
\end{align}
At high $P$, the dominating term in \eqref{eq:sem_asym_poutind} is the term
with $t = t_1
=\left\lceil\frac{BR}{I_{\mathcal X}(\beta)}\right\rceil-1 $. Therefore,
\begin{equation}
\Pout(P, R) \doteq \mathcal{K}_{\beta} P^{-md_{\beta}(R)},
\end{equation}
where
\begin{equation}
d_{\beta}(R) = B-t_1 = 1+\left\lfloor B\left(1-\frac{R}{I_{\mathcal
    X}(\beta)}\right) \right\rfloor.
\end{equation}
This concludes the proof of the proposition.
\end{proof}
Finally, from \eqref{eq:upbound_tw_outprob} and Proposition
\ref{prop:asym_poutind}, we have
\begin{equation}
\Pout(\pst^{\rm tw}(\boldgamma), P, R) \, \dot{\leq} \,
\mathcal{K}_{\beta}P^{-md_{\beta}(R)},
\end{equation}
as required by the Proposition.
\end{proof}

\newpage

\section{Proof of Theorem \ref{le:ref_sol}}
\label{app:ref_sol}
\begin{proof}
Similar to Theorem \ref{le:trunc_wfill_sol}, a solution
$\pst^{\rm ref}(\boldgamma)$ to the optimization problem given in
\eqref{eq:ref_prob} satisfies the KKT conditions  \cite{BoydVandenberghe2004} for the following problem:
\begin{equation}
\label{eq:simp_ref_prob}
\left\{\begin{array}{ll} {\rm Minimize} & - \sum_{b=1}^B I^{\rm ref}(p_b \gamma_b)\\
{\rm Subject \ to} & \sum_{b=1}^B p_b \leq BP\\
&p_b \leq \frac{\beta}{\gamma_b}, b=1, \ldots, B\\
& p_b \geq 0, b=1, \ldots, B\\
\end{array}\right.
\end{equation}
Therefore, $\mathbf{\pref}(\boldgamma)$ satisfies
\begin{align}
\label{eq:ref_cond1}\nu &\geq 0\\
\label{eq:ref_cond2}\sum_{b=1}^B \pref_b &\leq BP\\
\label{eq:ref_cond3}\nu\left(\sum_{b=1}^B \pref_b -BP\right)& =0\\
\label{eq:ref_cond4}\lambda_b &\geq 0, &b=1, \ldots, B\\
\label{eq:ref_cond5}\pref_b &\geq 0, &b=1, \ldots, B\\
\label{eq:ref_cond6}\lambda_b \pref_b &= 0, & b= 1, \ldots, B\\
\label{eq:ref_cond7}\tau_b &\geq 0, &b=1, \ldots, B\\
\label{eq:ref_cond8}\pref_b &\leq \frac{\beta}{\gamma_b}, &b=1, \ldots, B\\
\label{eq:ref_cond9}\tau_b \left(\pref_b- \frac{\beta}{\gamma_b}\right)& = 0, & b=1, \ldots, B
\end{align}
and
\begin{equation}
\label{eq:ref_cond10}
\left\{\begin{array}{lll}
-\frac{\gamma_b \log_2 e}{1+\pref_b \gamma_b} - \lambda_b + \tau_b + \nu &= 0,
&{\rm if\ } \pref_b < \frac{\alpha}{\gamma_b}\\
-\frac{\kappa \log_2 e}{\pref_b} - \lambda_b + \tau_b + \nu &= 0, &{\rm if \ }
\frac{\alpha}{\gamma_b} < \pref_b \leq \frac{\beta}{\gamma_b}\\
\frac{\gamma_b \log_2 e}{1+ \alpha} \geq -\lambda_b +\tau_b +\nu
 &\geq  \frac{\gamma_b\kappa \log_2 e}{\alpha}, &{\rm if \ } \pref_b = \frac{\alpha}{\gamma_b}
\end{array}\right.
\end{equation}
for $b= 1, \ldots, B$.

For any $b$, consider the following cases
\begin{itemize}
\item If $\lambda_b > 0$, then from \eqref{eq:ref_cond6},
  \eqref{eq:ref_cond9}, we have $\pref_b = \tau_b = 0$. In this case, condition
  \eqref{eq:ref_cond10} is satisfied only if $\nu > \gamma_b \log_2 e$.
\item If $\lambda_b = 0$ and $\tau_b > 0$, then from \eqref{eq:ref_cond9},
  $\pref_b = \frac{\beta}{\gamma_b}$. Therefore, from \eqref{eq:ref_cond10},
  $\nu < \frac{\kappa \gamma_b \log_2 e}{\beta}$.
\item If $\lambda_b = \tau_b = 0$, from \eqref{eq:ref_cond10}, we have
\begin{itemize}
\item[+] $\pref_b = \frac{\log_2 e}{\nu} - \frac{1}{\gamma_b}$ when $0 \leq
  \pref_b <  \frac{\alpha}{\gamma_b}$ or equivalently when $
  \frac{\gamma_b\log_2 e}{1+\alpha} < \nu \leq \gamma_b \log_2 e$.
\item[+] $\pref_b = \frac{\kappa \log_2 e}{\nu}$ if $\frac{\alpha}{\gamma_b} <
  \pref_b \leq \frac{\beta}{\gamma_b} \Leftrightarrow \frac{\kappa \log_2 e
  \gamma_b}{\beta} \leq \nu < \frac{\kappa \gamma_b \log_2 e }{\alpha}$.
\item[+] $\pref_b = \frac{\alpha}{\gamma_b}$ if $\frac{\gamma_b \kappa \log_2
  e}{\alpha} \leq \nu \leq \frac{\gamma_b \log_2 e}{1+\alpha}$.
\end{itemize}
\end{itemize}

Therefore, for any choice of $\nu$, we have
\begin{equation}
\label{eq:ref_sol_withnu}
\pref_b = \begin{cases}
\frac{\beta}{\gamma_b}, &\nu < \frac{\kappa \gamma_b \log_2 e}{\beta}\\
\frac{\kappa \log_2 e}{\nu}, & \frac{\kappa\gamma_b\log_2 e }{\beta}\leq \nu <
\frac{\kappa \gamma_b \log_2 e}{\alpha}\\
\frac{\alpha}{\gamma_b}, &\frac{\kappa \gamma_b \log_2 e}{\alpha} \leq \nu
\leq \frac{\gamma_b\log_2 e}{1+\alpha}\\
\frac{\log_2 e}{\nu}-\frac{1}{\gamma_b}, & \frac{\gamma_b \log_2 e}{1+\alpha}
< \nu \leq \gamma_b \log_2 e\\
0, &{\rm otherwise.}
\end{cases}
\end{equation}
We are left to choose $\nu \geq 0$ such that conditions
\eqref{eq:ref_cond1}--\eqref{eq:ref_cond3} are satisfied. If $\nu = 0$, then
from \eqref{eq:ref_sol_withnu}, $\pref_b= \frac{\beta}{\gamma_b}, ~b=1, \ldots,
B$. Furthermore, from \eqref{eq:ref_cond2}, $\nu= 0$ is valid only if
\begin{equation}
\sum_{b=1}^B \frac{\beta}{\gamma_b} \leq BP.
\end{equation}
If $\sum_{b=1}^B \frac{\beta}{\gamma_b}> BP$, then $\nu > 0$. Therefore, from
\eqref{eq:ref_cond3}, $\nu$ is chosen such that
\begin{equation}
\sum_{b=1}^B \pref_b = BP.
\end{equation}
Therefore, by denoting $\eta = \frac{\log_2 e}{\nu}$, we obtain
$\mathbf{\pref}(\boldgamma)$ as defined in the Theorem.
\end{proof}


\newpage

\section{Proof of Theorem \ref{theo:opt_longterm}}
\label{app:long_term_proof}
We first consider the following Proposition, which is a generalization of the
result in \cite{CaireTariccoBiglieri1999} to channels with discrete inputs.
\begin{proposition}
\label{prop:long_sol_with_w}
The solution of \eqref{eq:long_prob} has the following form
\begin{equation}
\label{eq:long_sol_with_w}
\pltopt(\boldgamma) = \begin{cases}
\boldlp^{\rm opt}(\boldgamma), &{\rm with \ probability\ }  \hat{w}(\boldgamma)\\
0, &{\rm with \  probability\ } 1-\hat{w}(\boldgamma),
\end{cases}
\end{equation}
where $\boldlp^{\rm opt}$ is the solution to the problem in
\eqref{eq:opt_min_pow_prob} and $\hat{w}(\boldgamma)$ is the
solution of
\begin{equation}
\label{eq:max_wgam_prob}
\left\{\begin{array}{ll}
{\rm Maximize} &\expectation{w(\boldgamma)}{} \\
{\rm Subject \ to} & 0 \leq w(\boldgamma) \leq 1\\
 &\expectation{\mean{\boldlp^{\rm opt}(\boldgamma)} w(\boldgamma)}{} \leq P.
\end{array}\right.
\end{equation}
\end{proposition}

\begin{proof}
From \eqref{eq:long_sol_with_w} and \eqref{eq:max_wgam_prob}, we have
\begin{equation}
\expectation{\mean{\pltopt(\boldgamma)}}{} = \expectation{\mean{\boldlp^{\rm
    opt}(\boldgamma)} \hat{w}(\boldgamma)}{} \leq P,
\end{equation}
which shows that $\pltopt(\boldgamma)$ satisfies the long-term power
constraint. We need to prove that
\begin{equation}
\label{eq:long_outprob_withw}
\Pout(\pltopt(\boldgamma), P, R) = 1 - \expectation{\hat{w}(\boldgamma)}{} \leq
\Pout(\mathbf{p}(\boldgamma), P, R),
\end{equation}
where $\mathbf{p}(\boldgamma)$ is an arbitrary power allocation
scheme satisfying the long-term power constraint
$\expectation{\mean{\mathbf{p}(\boldgamma)}}{} \leq P$.

Given a channel realization $\boldgamma$, define the region
\begin{equation}
\mathcal{A}(\boldgamma, R) \triangleq \left\{ \mathbf{p} \in \mathbb{R}_+^B:
\frac{1}{B}\sum_{b=1}^B \IX(p_b \gamma_b) \geq R \right\},
\end{equation}
and
\begin{equation}
\label{eq:def_w} w(\boldgamma) \triangleq \Pr\left(\mathbf{p}(\boldgamma) \in \mathcal{A}(\boldgamma, R)\right).
\end{equation}
Since $\mathcal{A}(\boldgamma, R)$ is the power allocation region that does not
cause outages, the outage probability given a channel realization $\boldgamma$
is $1-w(\boldgamma)$, and the overall outage probability is given by
\begin{equation}
\Pout(\mathbf{p}(\boldgamma), P, R) = 1- \expectation{w(\boldgamma)}{}.
\end{equation}
We now prove that $w(\boldgamma)$ satisfies the constraints of the
problem given in \eqref{eq:max_wgam_prob}.  By definition \eqref{eq:def_w} we
have that $0 \leq w(\boldgamma)\leq 1$. Furthermore, since $\boldlp^{\rm
  opt}(\boldgamma)$ is a solution to \eqref{eq:opt_min_pow_prob}, we have
\begin{equation}
\forall~ \mathbf{p}(\boldgamma) \in \mathcal{A}(\boldgamma, R), \;
\mean{\boldlp^{\rm opt}(\boldgamma)}\leq \mean{\mathbf{p}(\boldgamma)}.
\end{equation}
Therefore, conditioned on $\boldgamma$, the expectation of
$\mean{\mathbf{p}(\boldgamma)}$ over the distribution of
$\mathbf{p}(\boldgamma)$ can be lower bounded as follows.
\begin{align}
\expectation{\mean{\mathbf{p}(\boldgamma)}|\boldgamma}{\mathbf{p}(\boldgamma) \in \mathbb{R}_+^B}
&\geq\expectation{\mean{\mathbf{p}(\boldgamma)}|\boldgamma}{\mathbf{p}(\boldgamma) \in
  \mathcal{A}(\boldgamma, R)}\\
& \geq \mean{\boldlp^{\rm
    opt}(\boldgamma)} \Pr(\mathbf{p}(\boldgamma)\in
\mathcal{A}(\boldgamma, R))\\
&= \mean{\boldlp^{\rm opt}(\boldgamma)}w(\boldgamma).
\end{align}
Thus, since $\expectation{\mean{\mathbf{p}(\boldgamma)}}{} =
 \expectation{\expectation{\mean{\mathbf{p}(\boldgamma)}|\boldgamma}{\mathbf{p}(\boldgamma)
 \in \mathbb{R}_+^B}}{\boldgamma \in \mathbb{R}_+^B} \leq P$,
 we have
\begin{equation}
\expectation{\mean{\boldlp^{\rm opt}(\boldgamma)}w(\boldgamma)}{} \leq P.
\end{equation}
As a result, $w(\boldgamma)$ satisfies the constraints in \eqref{eq:max_wgam_prob}, and thus,
\[
\expectation{\hat{w}(\boldgamma)}{} \geq \expectation{w(\boldgamma)}{}.
\]

Therefore, we finally have
\begin{equation}
\Pout(\pltopt, P, R ) = 1- \expectation{\hat{w}(\boldgamma)}{}\leq 1 -
\expectation{w(\boldgamma)}{} = \Pout(\mathbf{p}, P, R)
\end{equation}
for any arbitrary $\mathbf{p}(\boldgamma)$, which shows that
$\pltopt(\boldgamma)$ is a solution to the problem given in \eqref{eq:long_prob}.
\end{proof}

We have the following Proposition, which gives the solution to
the problem given in \eqref{eq:max_wgam_prob}.
\begin{proposition}
\label{prop:sol_max_w} Suppose that $\gamma_b$ follows a continuous probability
density function. Then, a solution to the problem in \eqref{eq:max_wgam_prob}
is given by
\begin{equation}
\hat{w}(\boldgamma) = \begin{cases}
1, &{\rm if \ } \mean{\boldlp^{\rm opt}(\boldgamma)} \leq s\\
0, &{\rm otherwise,}
\end{cases}
\end{equation}
where $s$ satisfies
\begin{equation}
\label{eq:find_s}
\left\{
\begin{array}{lll}
s&= \infty, & {\rm if}~\lim_{s \to
      \infty}P(s) \leq P\\
P(s) &= P, &{\rm otherwise,}
\end{array} \right.
\end{equation}
and
\[
P(s)\triangleq \expectation{\mean{\boldlp^{\rm opt}(\boldgamma)}\hat{w}(\boldgamma)}{}.
\]
\end{proposition}

\begin{proof}
If $\lim_{s \to \infty} P(s) \leq P$, $\hat{w}(\boldgamma) = 1$ (which
corresponds to $s= \infty$) is certainly a solution to the problem.

Consider the case when $\lim_{s \to \infty} P(s) > P $.
 We first prove the existence of an $s$ satisfying \eqref{eq:find_s}.
Denoting $f_{\boldgamma}(\boldgamma)$ as the pdf of $\boldgamma$,
we can write $P(s)$ as
\begin{equation}
\label{eq:def_Ps_with_Rs} P(s) = \int_{\mathcal{R}(s)} \mean{\boldlp^{\rm
opt}(\boldgamma)} f_{\boldgamma}(\boldgamma)d\boldgamma,
\end{equation}
where $\mathcal{R}(s)$ is defined in \eqref{eq:def_Rs}. For all $s_0 >
s, \mathcal{R}(s) \subset \mathcal{R}(s_0)$. Therefore, from
\eqref{eq:def_Ps_with_Rs}, $P(s)$ is an
increasing function of $s$.
Due to the continuity of the fading statistics and
of the mutual information curve, $\mean{\boldlp^{\rm opt}(\boldgamma)}$ is a
continuous function of $\boldgamma$.

Without loss of generality, assume that $\gamma_1 \geq \ldots \geq
\gamma_B$. We first prove that $\lp_1^{\rm opt}> 0$. This is in
fact the case since for any power allocation scheme
$\boldlp(\boldgamma)$ such that $\lp_1(\boldgamma)= 0$, and
$\lp_k(\boldgamma)> 0$ satisfying the rate constraint, the power
  allocation scheme $\boldlp'(\boldgamma)$ with
\begin{equation}
\lp'_b(\boldgamma)= \begin{cases}
\lp_k(\boldgamma)\frac{\gamma_k}{\gamma_1}, &b=1\\
0, &b=k\\
\lp_b(\boldgamma), &{\rm otherwise,}
\end{cases}
\end{equation}
which has $\mean{\boldlp'(\boldgamma)} <
\mean{\boldlp(\boldgamma)}$, also satisfies the rate constraints.

Now assume that ${\boldgamma' }$ satisfies $\gamma_1' > \gamma_1, \gamma_b'=
\gamma_b, \, b=2, \ldots, B$. Consider the power allocation scheme
$\boldlp(\boldgamma')$ satisfying
\begin{equation}
\label{eq:proof_dec_lpopt}
\lp_b(\boldgamma') =\begin{cases}
 \lp_1^{\rm opt}(\boldgamma) \frac{\gamma_1}{\gamma_1'}, &b=1\\
 \lp_b^{\rm opt}(\boldgamma), &{\rm otherwise}.
\end{cases}
\end{equation}
Obviously,
\[
\sum_{b=1}^B \IX(\lp_b(\boldgamma')\gamma_b') = \sum_{b=1}^B \IX(\lp^{\rm
  opt}_b \gamma_b) \geq BR.
\]
Therefore, $\mean{\boldlp^{\rm opt}(\boldgamma')} \leq
\mean{\boldlp(\boldgamma')} < \mean{\boldlp^{\rm
opt}(\boldgamma)}$ (since $\lp_1^{\rm opt}(\boldgamma) > 0$). This
proves that $\mean{\boldlp^{\rm opt}(\boldgamma)}$ is a  strictly
decreasing function of $\gamma_1$ for any fixed $\gamma_2, \ldots,
\gamma_B$.

Due to the aforementioned monotonity and continuity of
$\mean{\boldlp^{\rm opt}(\boldgamma)}$, given $\gamma_2, \ldots,
\gamma_B$, there exists a unique $\gamma_1(s, \gamma_2, \ldots,
\gamma_B)$ such that $\mean{\boldlp^{\rm opt}(\boldgamma)}= s$ for
any $s > 0$, and we can rewrite the region $\mathcal{R}(s)$ as
\begin{equation}
\mathcal{R}(s)= \{\boldgamma \in \mathbb{R}_+^B: \gamma_1 \geq \gamma_1(s,
\gamma_2, \ldots, \gamma_B)\}.
\end{equation}
Additionally, for all $\epsilon$, there exists a $\delta(\epsilon)$ such that
\begin{align}
\gamma_1(s+\epsilon, \gamma_2, \ldots, \gamma_B) &= \gamma_1(s, \gamma_2, \ldots,
\gamma_B) + \delta(\epsilon)\\
\lim_{\epsilon \to 0} \delta(\epsilon) &= 0\\
\epsilon \delta(\epsilon) &\leq 0.
\end{align}
Therefore, denoting $\gamma_s \triangleq \gamma_1(s, \gamma_2, \ldots,
\gamma_B)$, we have
\begin{align}
\lim_{\epsilon \to 0} P(s+\epsilon) &= \lim_{\epsilon \to 0}
\int_{\mathcal{R}(s+\epsilon)}\mean{\boldlp^{\rm
    opt}(\boldgamma)}f_{\boldgamma}(\boldgamma)d\boldgamma  \\
&= \lim_{\epsilon \to 0} \int_{\gamma_2,
  \ldots, \gamma_B} \left(\int_{\gamma_s+\delta(\epsilon)}^{\infty} \mean{\boldlp^{\rm
    opt}(\boldgamma)}f_{\boldgamma}(\boldgamma)d\gamma_1 \right) d\gamma_2 \ldots
d\gamma_B\\
&= \int_{\gamma_2,
  \ldots, \gamma_B} \left(\lim_{\delta(\epsilon) \to 0} \int_{\gamma_s+\delta(\epsilon)}^{\infty} \mean{\boldlp^{\rm
    opt}(\boldgamma)}f_{\boldgamma}(\boldgamma)d\gamma_1 \right) d\gamma_2 \ldots
d\gamma_B\\
    &= \int_{\gamma_2,
  \ldots, \gamma_B} \left( \int_{\gamma_s}^{\infty} \mean{\boldlp^{\rm
    opt}(\boldgamma)}f_{\boldgamma}(\boldgamma)d\gamma_1 \right) d\gamma_2 \ldots
d\gamma_B\\
&=P(s).
\end{align}
Thus, $P(s)$ is an continuously increasing function of $s$, which proves that
there exists an $s$ satisfying $P(s) = P$ since $\lim_{s \to \infty} P(s) > P$.

On the other hand, for any $w(\boldgamma)$, we have
\begin{align}
\expectation{\mean{\boldlp^{\rm opt}(\boldgamma)}w(\boldgamma)}{}-P &=
\expectation{\mean{\boldlp^{\rm
    opt}(\boldgamma)}w(\boldgamma)}{}-  \expectation{\mean{\boldlp^{\rm
    opt}(\boldgamma)}\hat{w}(\boldgamma)}{}\\
\label{eq:explain_a} &= \int_{\mathbb{R}_+^B \setminus
\mathcal{R}(s)} w(\boldgamma) \mean{\boldlp^{\rm
    opt}(\boldgamma)}dF_{\boldgamma}(\boldgamma) \\
    & \hspace{10mm}- \int_{\mathcal{R}(s)}
    (1-w(\boldgamma))\mean{\boldlp^{\rm opt}(\boldgamma)} dF_{\boldgamma}(\boldgamma)\\
\label{eq:explain_b}&\geq s\left(\int_{\mathbb{R}_+^B \setminus
\mathcal{R}(s)}
    w(\boldgamma)dF_{\boldgamma}(\boldgamma)-
    \int_{\mathcal{R}(s)}(1-w(\boldgamma))dF_{\boldgamma}(\boldgamma)\right)\\
\label{eq:explain_c}&= s\left(\expectation{w(\boldgamma)}{} - \expectation{\hat{w}(\boldgamma)}{}\right),
\end{align}
where \eqref{eq:explain_a} and \eqref{eq:explain_c} are due to
\begin{equation}
\hat{w}(\boldgamma) = \begin{cases}
1, &{\rm if\ } \boldgamma \in \mathcal{R}(s)\\
0, &{\rm otherwise,}
\end{cases}
\end{equation}
and \eqref{eq:explain_b} is obtained using the following bounds
\begin{align}
\mean{\boldlp^{\rm opt}(\boldgamma)} &\leq s, \hspace{5mm} {\rm if
\ } \boldgamma \in
\mathcal{R}(s)\\
\mean{\boldlp^{\rm opt}(\boldgamma)} & > s, \hspace{5mm} {\rm if \
} \boldgamma \notin \mathcal{R}(s).
\end{align}

Therefore, for all $w(\boldgamma)$, $\expectation{w(\boldgamma)}{} \geq
\expectation{\hat{w}(\boldgamma)}{}$ implies $\expectation{\mean{\boldlp^{\rm
      opt}(\boldgamma)}w(\boldgamma)}{} > P$, which violates the problem constraint.
Thus, $\hat{w}(\boldgamma)$ is a solution to the problem. This concludes the
proof of the proposition.
\end{proof}

The proof of the Theorem is obtained by  applying  Propositions
\ref{prop:long_sol_with_w} and \ref{prop:sol_max_w}.

\newpage
\section{Proof of Theorem \ref{lem:opt_minpow_sol}}
\label{app:opt_minpow_sol}
\begin{proof}
Since $\mean{\boldlp}, -\sum_{b=1}^B \IX(\lp\gamma_b), -\lp_b$ are convex
functions of $\boldlp (\boldgamma)$, applying the KKT conditions
\cite{BoydVandenberghe2004} and note the fact that
\cite{GuoShamaiVerdu2005}
\begin{equation}
\frac{d}{d \rho} \IX(\rho) = \frac{1}{\log 2}\MMSE(\rho),
\end{equation}
the solution $\boldlp^{\rm opt}(\boldgamma)$ of \eqref{eq:opt_min_pow_prob}
satisfies the following conditions
\begin{align}
\label{eq:opt_minpow_sol1}\nu &\geq 0\\
\label{eq:opt_minpow_sol2}\nu\left(BR - \sum_{b=1}^B \IX(\lp^{\rm opt}_b \gamma_b) \right)&= 0\\
\label{eq:opt_minpow_sol3}BR -\sum_{b=1}^B \IX(\lp^{\rm opt}_b \gamma_b) &\leq 0\\
\label{eq:opt_minpow_sol4}\lambda_b &\geq 0, &b=1, \ldots, B\\
\label{eq:opt_minpow_sol5}-\lp^{\rm opt}_b &\leq 0, &b=1, \ldots, B\\
\label{eq:opt_minpow_sol6}\lambda_b \lp^{\rm opt}_b &= 0, &b=1, \ldots, B\\
\label{eq:opt_minpow_sol7}1 -\frac{1}{\log 2}\nu \gamma_b \MMSE(\lp^{\rm opt}_b
\gamma_b)-\lambda_b &= 0, & b=1, \ldots, B
\end{align}
where $\nu, \lambda_b, b=1, \ldots, B$ are the Lagrangian multipliers. Letting
$\eta = \frac{\nu}{\log 2}$, for any $b$, we have
\begin{itemize}
\item If $\lambda_b > 0$, then from \eqref{eq:opt_minpow_sol6}, $\lp^{\rm opt}_b
= 0$ and thus \eqref{eq:opt_minpow_sol7} requires $\eta < \frac{1}{\gamma_b}$.
\item If $\lambda_b = 0$, then from \eqref{eq:opt_minpow_sol7},
\begin{equation}\
\lp^{\rm opt}_b = \frac{1}{\gamma_b}\MMSE^{-1}\left(\frac{1}{\eta
\gamma_b}\right)
\end{equation}
and $\eta \geq \frac{1}{\gamma_b}$ since $\MMSE(\rho) \leq 1$.
\end{itemize}
Therefore, with any choice of $\eta$, we have
\begin{align}
\label{eq:app_sol_minpow} \lp^{\rm opt}_b &= \left\{\begin{array}{ll}
\frac{1}{\gamma_b}\MMSE^{-1}\left(\frac{1}{\eta \gamma_b}\right), &\gamma_b
\geq \frac{1}{\eta} \\
0, &{\rm otherwise}
\end{array}\right.\\
&= \frac{1}{\gamma_b}\MMSE^{-1}\left( \min\left\{1, \frac{1}{\eta \gamma_b}\right\}\right),
\end{align}
for $b=1, \ldots, B$.
We are left to choose $\eta \geq 0$ such that \eqref{eq:opt_minpow_sol2}
and \eqref{eq:opt_minpow_sol3} are satisfied. From \eqref{eq:app_sol_minpow},
\eqref{eq:opt_minpow_sol3} is not satisfied if $\eta = 0$. Therefore, from
\eqref{eq:opt_minpow_sol2}, $\eta$ is chosen such that
\begin{equation}
\label{eq:opt_minpow_eta} \sum_{b=1}^B \IX(\lp^{\rm opt}_b \gamma_b) = BR
\end{equation}
as required by the Theorem.
\end{proof}

\newpage

\section{Asymptotic Analysis of Power Allocation for Long-Term Constraints}
\label{se:long_asym}

\paragraph*{Proof of Proposition \ref{prop:diff_Ps}}

From the definition of differentiation, we have
\begin{equation}
\label{eq:app_def_diff_Ps}
\frac{d}{ds}P(s) = \lim_{a \downarrow 1} \frac{P(as)-P(s)}{as - s},
\end{equation}
where
\begin{align}
P(as) &= \expectation{\mean{\boldlp(\boldgamma)}}{\boldgamma \in \mathcal{R}(as)}\\
&= \expectation{\mean{\boldlp(\boldgamma)}}{\boldgamma \in \mathcal{R}(s)} +
\expectation{\mean{\boldlp(\boldgamma)}}{\boldgamma \in \mathcal{R}(as)
  \setminus \mathcal{R}(s)}\\
\label{eq:app_Pas} &= P(s)+ \expectation{\mean{\boldlp(\boldgamma)}}
{\boldgamma \in \mathcal{R}(as) \setminus \mathcal{R}(s)}.
\end{align}
Note that $\forall \boldgamma \in \mathcal{R}(s)$, we have that
$\mean{\boldlp(\boldgamma)} \leq s$. Therefore, since $a>1$,
$\mean{\boldlp(\boldgamma)} < as$, which implies that $\boldgamma \in
\mathcal{R}(as)$ and thus, $\mathcal{R}(s) \subset \mathcal{R}(as)$.
Now, let $f_{\boldgamma}(\boldgamma)$  be the pdf of the
$\boldgamma$.  Since $\forall \boldgamma \in \mathcal{R}(as),
\mean{\boldlp(\boldgamma)} \leq as$, we have
\begin{align}
 \expectation{\mean{\boldlp(\boldgamma)}}{\boldgamma \in \mathcal{R}(as)
 \setminus \mathcal{R}(s)} &= \int_{\boldgamma \in \mathcal{R}(as)
 \setminus \mathcal{R}(s)} \mean{\boldlp(\boldgamma)}
 f_{\boldgamma}(\boldgamma) d\boldgamma\\
&\leq as \int_{\boldgamma \in \mathcal{R}(as) \setminus
\mathcal{R}(s)} f_{\boldgamma}(\boldgamma) d\boldgamma\\
&= as \left[\Pr\left(\boldgamma \notin \mathcal{R}(s)\right)
- \Pr\left(\boldgamma \notin \mathcal{R}(as)\right)\right].
\end{align}
From the assumption in \eqref{eq:app_short_long_asym}, and noting
that $\Pout(\plt(\boldgamma), P(s),R) = \Pr(\boldgamma \notin
\mathcal{R}(s)) \doteq \mathcal{K}s^{-d(R)}$, we have
\begin{equation}
\label{eq:app_upbound_Pas}
\expectation{\mean{\boldlp(\boldgamma)}}{\boldgamma \in \mathcal{R}(as)
\setminus \mathcal{R}(s)} ~\dot{\leq} ~ as \mathcal{K} \left(s^{-d(R)}- (as)^{-d(R)}\right).
\end{equation}
On the other hand, since $\forall \boldgamma \in \mathcal{R}(as)
\setminus \mathcal{R}(s), \mean{\boldlp(\boldgamma)} > s$, by
similar arguments,
\begin{equation}
\label{eq:app_lowbound_Pas}
\expectation{\mean{\boldlp(\boldgamma)}}{\boldgamma \in
\mathcal{R}(as) \setminus \mathcal{R}(s)} ~\dot{\geq}~ s
\mathcal{K} \left(s^{-d(R)}- (as)^{-d(R)}\right).
\end{equation}
Therefore, from \eqref{eq:app_def_diff_Ps}, \eqref{eq:app_Pas},
\eqref{eq:app_upbound_Pas}, \eqref{eq:app_lowbound_Pas}, we have
\begin{equation}
\lim_{a \downarrow 1}\frac{ \mathcal{K}  s^{-d(R)}  \left(1 -
a^{-d(R)}\right)}{a-1} ~ \dot{\leq}~ \frac{d}{ds}P(s) ~\dot{\leq}~
\lim_{a \downarrow 1}\frac{a \mathcal{K}  s^{-d(R)} \left(1 -
a^{-d(R)}\right)}{a-1}.
\end{equation}
Since
\begin{equation}
\lim_{a \downarrow 1}\frac{a \mathcal{K}  s^{-d(R)}  \left(1 -
a^{-d(R)}\right)}{a-1}  =  \lim_{a \downarrow 1}\frac{\mathcal{K}
s^{-d(R)} \left(1 - a^{-d(R)}\right)}{a-1} = \mathcal{K}d(R)
s^{-d(R)},
\end{equation}
we have that
\begin{equation}
\frac{d}{ds} P(s) \doteq \mathcal{K} d(R) s^{-d(R)}
\end{equation}
which concludes the proof. \endproof


\paragraph*{Proof of Theorem \ref{theo:long_term_asymp}}
We begin with the first part of the theorem, i.e.  $\lim_{s \to
\infty} P(s) = P_{\rm th} < \infty$ when $d(R)> 1$. From
Proposition \ref{prop:diff_Ps}, we have
\begin{equation}
\lim_{s \to \infty} \left(\frac{d}{ds}P(s) \right) s^{d(R)} = \mathcal{K} d(R).
\end{equation}
Therefore, for any $\epsilon > 0$, there exists a finite $s_1$
such that for all $s> s_1$,
\begin{equation}
\left(\frac{d}{ds}P(s) \right) s^{d(R)} < \mathcal{K} d(R)+ \epsilon,
\end{equation}
or equivalently, for all $s> s_1$,
\begin{equation}
\frac{d}{ds}P(s) < (\mathcal{K} d(R) + \epsilon )s^{-d(R)}.
\end{equation}
Thus,
\begin{align}
\lim_{ s \to \infty} P(s) &= P(s_1)+ \lim_{s \to \infty} \int_{s_1}^{s} \left(\frac{d}{dt}P(t)\right)dt\\
&< P(s_1)+\lim_{s \to \infty} \int_{s_1}^{s} (\mathcal{K} d(R)+\epsilon) t^{-d(R)} dt\\
&= P(s_1)+ \lim_{s \to \infty} \frac{\left(\mathcal{K} d(R)+
\epsilon)(s^{1-d(R)} - s_1^{1-d(R)}\right)}{1-d(R)}, \label{eq:limps}
\end{align}
which gives
\begin{align}
\lim_{ s \to \infty} P(s) &< P(s_1)+ \frac{(\mathcal{K} d(R)+ \epsilon) s_1^{1-d(R)}}{d(R)-1}\\
&\triangleq P_{\rm th} < \infty,
\end{align}
when $d(R) > 1$ as required.
Furthermore, from the definition of the long-term exponent,
\begin{align}
d_{\rm lt}(R) &= \lim_{P(s) \to \infty} \frac{-\log \Pout(\plt(\boldgamma), P(s),R)}{\log P(s)}\\
&= \lim_{s \to \infty} \frac{-\log \Pout(\plt(\boldgamma), P(s),R)}{\log P(s)}\\
&= \lim_{s \to \infty} \frac{-\log \left(\mathcal{K} s^{-d(R)}\right)}{\log P(s)}\\
\label{eq:simp_dlt}&= \lim_{s \to \infty} \frac{d(R)\log s}{\log P(s)}.
\end{align}
Therefore
\begin{equation}
d_{\rm lt}(R) = \lim_{s \to \infty} \frac{d(R)\log(s)}{P_{\rm th}} = \infty
\end{equation}
if $d(R)>1$.

In the second part of the theorem, where we have $d(R) < 1$, then
from \eqref{eq:limps} we observe that $\lim_{s \to \infty}P(s) =
\infty$. Applying L'H\^opital's rule to \eqref{eq:simp_dlt}, we
obtain
\begin{equation}
d_{\rm lt}(R) = \lim_{s \to \infty}
\frac{d(R)\frac{P(s)}{s}}{\frac{d}{ds}P(s)}.
\end{equation}
Applying Proposition \ref{prop:diff_Ps}, we can further write $d_{\rm lt}(R)$ as
\begin{equation}
d_{\rm lt}(R) = \lim_{s \to \infty} \frac{P(s)}{\mathcal{K}s^{1-d(R)}}.
\end{equation}
Further applying L'H\^opital's rule and Propostion \ref{prop:diff_Ps} yields
\begin{align}
d_{\rm lt}(R) &= \lim_{s \to \infty} \frac{\frac{d}{ds}P(s)}{\mathcal{K}(1-d(R))s^{-d(R)}}\\
&=\frac{d(R)}{1-d(R)},
\end{align}
which completes the proof.
\endproof

\paragraph*{Proof of Proposition \ref{prop:dual_minpow_maxrate}}

From \cite{LozanoTulinoVerdu2006}, the optimal power allocation scheme for system
with short-term power constraint $s$ is given by
\begin{equation}
\popt_b(\boldgamma) = \frac{1}{\gamma_b}\MMSE^{-1}\left(\min\left\{1,
\frac{\eta}{\gamma_b}\right\}\right), \; b=1, \ldots, B,
\end{equation}
where $\eta$ is chosen such that the power constraint is satisfied,
\begin{equation}
\sum_{b=1}^B \popt_b(\boldgamma) = Bs.
\end{equation}

Transmission with rate $R$ and power allocation scheme
$\pst^{\rm opt}(\boldgamma)$ is in outage if and only if there is no $\eta$ satisfying
\begin{align}
\label{eq:out_cond1}\sum_{b=1}^B \IX\left(\MMSE^{-1}\left(\min\left\{1, \frac{1}
{\eta \gamma_b}\right\}\right)\right) &\geq BR\\
\label{eq:out_cond2}\sum_{b=1}^B \frac{1}{\gamma_b}\MMSE^{-1} \left(\min
\left\{1, \frac{1}{\eta \gamma_b} \right\} \right) &\leq Bs.
\end{align}
Similarly, from Lemma \ref{lem:opt_minpow_sol} and Theorem
  \ref{theo:opt_longterm}, transmission with power allocation scheme
  $\plt^{\rm opt}(\boldgamma)$ is also in outage if and only if there is no $\eta$
  satisfying \eqref{eq:out_cond1} and \eqref{eq:out_cond2} simultaneously.

Therefore,
\begin{equation}
\label{eq:dual_short_short}
\Pout(\plt(\boldgamma),P(s), R) = \Pout(\pst^{\rm opt}(\boldgamma),
s, R)
\end{equation}
as required by the Proposition.
\endproof
\newpage
\section{Proof of Theorem \ref{lem:tw_minpow_sol}}
\label{app:tw_minpow_sol}
\begin{proof}
Similar to Appendix \ref{app:tw_sol}, a solution to the problem given in
\eqref{eq:tw_min_pow_prob} is given by  solving
\begin{equation}
\label{eq:tw_minpow_prob_sim}
\left\{\begin{array}{ll} {\rm Minimize} &\mean{\boldlp(\boldgamma)}\\
{\rm Subject \ to} & \sum_{b=1}^B \log_2(1+\lp_b \gamma_b) \geq BR\\
&\lp_b \leq \frac{\beta}{\gamma_b}\\
& \lp_b \geq 0, \;\; b= 1, \ldots, B.
\end{array}\right.
\end{equation}
The problem given in \eqref{eq:tw_minpow_prob_sim} is a standard
convex optimization problem. Therefore, according to the KKT
conditions \cite{BoydVandenberghe2004}, a solution
$\boldlp^{\star}(\boldgamma)$ to the problem satisfies
\begin{align}
\label{eq:tw_minpow_cond1} \nu & \geq 0\\
\label{eq:tw_minpow_cond2}BR - \sum_{b=1}^B \log_2(1+ \lp^{\star}_b\gamma_b) &\leq 0\\
\label{eq:tw_minpow_cond3}\nu \left(BR - \sum_{b=1}^B \log_2(1+ \lp^{\star}_b\gamma_b )\right)& = 0\\
\label{eq:tw_minpow_cond4} \lambda_b &\geq 0, & b= 1, \ldots, B\\
\label{eq:tw_minpow_cond5} \lp^{\star}_b &\geq 0, &b=1, \ldots, B\\
\label{eq:tw_minpow_cond6} \lambda_b \lp^{\star}_b & = 0, & b= 1, \ldots, B\\
\label{eq:tw_minpow_cond7} \alpha_b & \geq 0, & b= 1, \ldots, B\\
\label{eq:tw_minpow_cond8} \lp^{\star}_b -\frac{\beta}{\gamma_b}& \leq 0,
&b=1, \ldots, B\\
\label{eq:tw_minpow_cond9} \alpha(\lp^{\star}_b - \frac{\beta}{\gamma_b})& =
0, & b= 1, \ldots, B\\
\label{eq:tw_minpow_cond10} 1- \nu \frac{\gamma_b \log_2 e}{1 + \lp^{\star}_b
  \gamma_b } -\lambda_b + \alpha_b &= 0, &b= 1, \ldots, B
\end{align}
where $\nu, \lambda_b, \alpha_b, b=1, \ldots, B$ are the Lagrangian
multipliers. For any $b$,
\begin{itemize}
\item If $\lambda_b > 0$, from \eqref{eq:tw_minpow_cond6} and
  \eqref{eq:tw_minpow_cond9}, we have $\lp^{\star}_b = \alpha_b =
  0$. Therefore, \eqref{eq:tw_minpow_cond10} requires $\nu < \frac{1}{\gamma_b
  \log_2 e}$.
\item If $\lambda_b = 0, \alpha_b > 0$, from \eqref{eq:tw_minpow_cond9}, we
  have $\lp^{\star}_b = \frac{\beta}{\gamma_b}$. Therefore,
  \eqref{eq:tw_minpow_cond10} requires $\nu > \frac{\beta+1}{\gamma_b \log_2
  e}$.
\item If $\lambda_b = \alpha_b = 0$, from \eqref{eq:tw_minpow_cond10}, we have
  $\lp^{\star}_b = \nu \log_2 e - \frac{1}{\gamma_b}$. Therefore,
  \eqref{eq:tw_minpow_cond4} and \eqref{eq:tw_minpow_cond8} require
  $\frac{1}{\gamma_b \log_2 e} \leq \nu \leq \frac{\beta+1}{\gamma_b \log_2
  e}$.
\end{itemize}

Therefore,  for any choice of $\nu$, we
have
\begin{align}
\lp^{\star}_b &= \begin{cases}
\frac{\beta}{\gamma_b}, &\nu >  \frac{\beta+1}{\gamma_b \log_2 e}\\
\nu \log_2 e- \frac{1}{\gamma_b}, &\frac{1}{\gamma_b \log_2 e} \leq \nu \leq
\frac{\beta+1}{\gamma_b \log_2 e}\\
0, &{\rm otherwise}
\end{cases}\\
\label{eq:tw_minpow_lpnu}&= \min\left\{\frac{\beta}{\gamma_b}, \left(\nu \log_2
e - \frac{1}{\gamma_b}
  \right)_+\right\}.
\end{align}
We are left to choose $\nu$ such that the  conditions in
\eqref{eq:tw_minpow_cond1}--\eqref{eq:tw_minpow_cond3} are satisfied. From
\eqref{eq:tw_minpow_lpnu}, \eqref{eq:tw_minpow_cond2} is not satisfied if $\nu =
0$. Therefore, from \eqref{eq:tw_minpow_cond3}, we choose $\nu$ such that
\begin{equation}
\sum_{b=1}^B \lp^{\star}_b = \sum_{b=1}^B \min\left\{\frac{\beta}{\gamma_b}, \left(\nu \log_2 e - \frac{1}{\gamma_b}
  \right)_+\right\} = BR.
\end{equation}
Finally, denoting $\eta = \nu \log_2 e$, we have
$\boldlp^{\star}(\boldgamma)$ as required by the Theorem.
\end{proof}

\newpage
\section{Proof of Proposition \ref{prop:dual_tw_minpow_maxrate}}
\label{app:dual_tw}
\begin{proof}
According to Theorem \ref{le:trunc_wfill_sol}, the truncated water-filling
scheme with short-term power constraint $s$ can be written as follows
\begin{equation}
\pst^{\rm tw}= f(\eta_{\rm st}, \gamma_b)
\end{equation}
with $\eta_{\rm st}$ chosen such that
\begin{equation}
\begin{cases}
\eta_{\rm st} = \infty &{\rm if \  } \sum_{b=1}^B \frac{1}{\gamma_b} \leq Bs\\
g(\eta, \boldgamma) = Bs &{\rm otherwise,}
\end{cases}
\end{equation}
where
\begin{align}
f(\eta, \boldgamma) &\triangleq  \min\left\{\frac{\beta}{\gamma_b}, \left(\eta-
\frac{1}{\gamma_b}\right)_+\right\},  \\
g(\eta, \boldgamma) &\triangleq \sum_{b=1}^B f(\eta, \gamma_b).
\end{align}

Similarly, from \eqref{eq:lp_tw}, $\lp^{\rm tw}_b = f(\eta_{\rm lt},
\gamma_b), \,b=1, \ldots, B$
with $\eta_{\rm lt}$ chosen such that
\begin{equation}
\label{eq:def_Ieta}
I(\eta_{\rm lt}, \boldgamma)   = BR,
\end{equation}
where $I(\eta, \boldgamma) \triangleq \sum_{b=1}^{B} \IX(f(\eta, \gamma_b)
\gamma_b)$. Transmission with power allocation scheme $\plt^{\rm tw}(\boldgamma)$ is
in outage if $g(\eta_{\rm lt}, \gamma_b) > Bs$.

Consider truncated water-filling schemes with $\beta$ chosen such
that $\IX(\beta) \geq R$. Consider a channel realization
$\boldgamma$, and, without loss of generality, suppose $\gamma_1
\geq \ldots \geq \gamma_B$. If transmission with power allocation
scheme $\pst^{\rm tw}(\boldgamma)$ is in outage then
\begin{align}
\eta_{\rm st} &< \frac{\beta+1}{\gamma_B}\\
g(\eta_{\rm st}, \boldgamma) &= Bs\\
\label{eq:Ieta_st}I(\eta_{\rm st}, \boldgamma) &< BR.
\end{align}
Noting that $I(\eta, \boldgamma)$ and $g(\eta, \boldgamma)$ are
increasing function of $\eta$ for $\eta <
\frac{\beta+1}{\gamma_B}$, from \eqref{eq:def_Ieta} and
\eqref{eq:Ieta_st}, we have $\eta_{\rm lt} > \eta_{\rm st}$ and
thus, $g(\eta_{\rm lt}, \boldgamma) > g(\eta_{\rm st}, \boldgamma)
= Bs$. Therefore, transmission with power allocation scheme
$\plt^{\rm tw}(\boldgamma)$ is also in outage.

By similar arguments, we also conclude that if transmission with
$\plt^{\rm tw}(\boldgamma)$ results in an outage event, then
transmission with $\pst^{\rm tw}(\boldgamma)$ also results in
outage.

Therefore,
\begin{equation}
\Pout(\plt^{\rm tw}(\boldgamma), P(s), R) = \Pout(\pst^{\rm tw}(\boldgamma), s,
R).
\end{equation}
\end{proof}
\newpage
\section{Proof of Theorem \ref{theo:ref_minpow}}
\label{app:ref_minpow_sol}
\begin{proof}
Similar to Theorem \ref{le:trunc_wfill_sol}, a solution
$\boldlp^{\star}(\boldgamma)$ to problem \eqref{eq:ref_min_pow_prob} satisfies
the KKT conditions \cite{BoydVandenberghe2004} for the following problem
\begin{equation}
\left\{\begin{array}{ll}
{\rm Minimize} & \mean{\boldlp(\boldgamma)}\\
{\rm Subject \ to}& \sum_{b=1}^B I^{\rm ref}(\lp_b \gamma_b) \geq BR\\
&\lp_b \leq \frac{\beta}{\gamma_b}, \;\;b=1, \ldots, B\\
&\lp_b \geq 0, \;\;b=1, \ldots, B.
\end{array}\right.
\end{equation}
Therefore, $\boldlp^{\star}(\boldgamma)$ satisfies
\begin{align}
\label{eq:ref_minpow_cond1}\nu &\geq 0\\
\label{eq:ref_minpow_cond2}BR - \sum_{b=1}^B I^{\rm ref}(\lp^{\star}_b \gamma_b) &\leq 0\\
\label{eq:ref_minpow_cond3} \nu \left(BR- \sum_{b=1}^B I^{\rm
  ref}(\lp^{\star}_b \gamma_b)\right)&= 0\\
\label{eq:ref_minpow_cond4}\lambda_b &\geq 0, &b=1, \ldots, B\\
\label{eq:ref_minpow_cond5}\lp^{\star}_b &\geq 0, &b=1, \ldots, B\\
\label{eq:ref_minpow_cond6} \lambda_b \lp^{\star}_b &= 0, &b=1, \ldots, B\\
\label{eq:ref_minpow_cond7} \tau_b &\geq 0, &b=1, \ldots, B\\
\label{eq:ref_minpow_cond8} \lp^{\star}_b -\frac{\beta}{\gamma_b} &\leq 0,
&b=1, \ldots, B\\
\label{eq:ref_minpow_cond9} \tau_b\left(\lp^{\star}_b
-\frac{\beta}{\gamma_b}\right) &= 0, &b=1, \ldots, B
\end{align}
and
\begin{equation}
\label{eq:ref_minpow_cond10}
\begin{cases}
1-\nu \frac{\gamma_b \log_2 e}{1+\lp^{\star}_b \gamma_b} - \lambda_b + \tau_b =
0, &{\rm if\ } \lp^{\star}_b < \frac{\alpha}{\gamma_b}\\
1- \nu \frac{\kappa \log_2 e}{\lp^{\star}_b} -\lambda_b + \tau_b = 0, &{\rm
  if \ } \frac{\alpha}{\gamma_b} < \lp^{\star}_b \leq \frac{\beta}{\gamma_b}\\
\nu \frac{\gamma_b \log_2 e}{1+ \tau} \geq 1 - \lambda_b +
\tau_b \geq \nu \frac{\gamma_b \kappa \log_2 e}{\alpha}, &{\rm if\ }
\lp^{\star}_b = \frac{\alpha}{\gamma_b}
\end{cases}
\end{equation}
for $b=1, \ldots, B$.

For any $b$, consider the following cases:
\begin{itemize}
\item If $\lambda_b > 0$, from \eqref{eq:ref_minpow_cond6} and
  \eqref{eq:ref_minpow_cond9}, we have $\lp^{\star}_b = \tau_b = 0$. In this
  case, condition  \eqref{eq:ref_minpow_cond10} is satisfied only if $\nu <
  \frac{1}{\gamma_b \log_2 e}$.
\item If $\lambda_b =0$ and $\tau_b > 0$, from  \eqref{eq:ref_minpow_cond9},
  $\lp^{\star}_b = \frac{\beta}{\gamma_b}$. Therefore, from
  \eqref{eq:ref_minpow_cond10}, $\nu > \frac{\beta}{\kappa \gamma_b \log_2
  e}$.
\item If $\lambda_b= \tau_b = 0$, from  \eqref{eq:ref_minpow_cond10}, we have
\begin{itemize}
\item[+] $\lp^{\star}_b = \nu \log_2 e - \frac{1}{\gamma_b} $ when
  $0 \leq \lp^{\star}_b < \frac{\alpha}{\gamma_b}$ or equivalently, when
  $\frac{1}{\gamma_b \log_2 e} \leq \nu <
  \frac{\alpha+1}{\gamma_b \log_2 e}$.
\item[+] $\lp^{\star}_b = \nu \kappa \log_2 e$ when $\frac{\alpha}{\gamma_b} <
  \lp^{\star}_b \leq \frac{\beta}{\gamma_b}$, or equivalently when
  $\frac{\alpha}{\kappa \gamma_b \log_2 e} < \nu \leq \frac{\beta}{\kappa
  \gamma_b \log_2 e}$.
\item[+] $\lp^{\star}_b = \frac{\alpha}{\gamma_b}$ when
  $\frac{1+\alpha}{\gamma_b \log_2 e} \leq \nu \leq
  \frac{\alpha}{\kappa\gamma_b\log_2 e}$.
\end{itemize}
\end{itemize}
Therefore, for any choice of $\nu$, we have

\begin{equation}
\label{eq:ref_minpow_sol_withnu}
\lp^{\star}_b = \begin{cases}
\frac{\beta}{\gamma_b}, &\nu > \frac{\beta}{\kappa \gamma_b \log_2 e}\\
\nu \kappa \log_2 e, & \frac{\alpha}{\kappa\gamma_b \log_2 e} < \nu \leq
\frac{\beta}{\kappa \gamma_b \log_2 e}\\
\frac{\alpha}{\gamma_b}, &\frac{1+\alpha}{\gamma_b \log_2 e} \leq \nu \leq
\frac{\alpha}{\kappa \gamma_b \log_2 e}\\
\nu \log_2 e -\frac{1}{\gamma_b}, & \frac{1}{\gamma_b \log_2 e} \leq \nu <
\frac{\alpha+1}{\gamma_b \log_2 e}\\
0, &{\rm otherwise.}
\end{cases}
\end{equation}
We are left to choose $\nu$ such that conditions
\eqref{eq:ref_minpow_cond1}-- \eqref{eq:ref_minpow_cond3} are
satisfied. From \eqref{eq:ref_minpow_cond10}, $\lp^{\star}_b= 0,
~b=1, \ldots, B$ if $\nu = 0$.  Therefore,
\eqref{eq:ref_minpow_cond2} requires that $\nu > 0$. Thus, from
\eqref{eq:ref_minpow_cond3}, we need to choose $\nu$ such that
\begin{equation}
\sum_{b=1}^B I^{\rm ref}(\lp^{\star}_b \gamma_b) = BR.
\end{equation}
Therefore, by denoting $\eta = \nu \log_2 e$, we obtained
$\boldlp^{\star}(\boldgamma)$ as defined in the Theorem.
\end{proof}

\newpage
\bibliographystyle{IEEEtran}
\bibliography{bib_database}


\newpage
\begin{table}[htbp]
  \centering
  \caption{Parameters $\rho_0,\kappa,a$ and $\alpha$ for the refined power allocation scheme.}

  \begin{tabular}{@{}ccccccccc@{}}
    \toprule
   &  \multicolumn{8}{c}{Modulation Scheme} \\ \cmidrule(l){2-9}
   &  \multicolumn{2}{c}{QPSK}  & \multicolumn{2}{c}{$8$-PSK}  & \multicolumn{2}{c}{$16$-QAM} & \multicolumn{2}{c}{$64$-QAM}\\\cmidrule(l){2-9}
   & CM & BICM & CM & BICM & CM & BICM & CM & BICM\\
    $\rho_0$& $3$ & $3$ & $7$ & $7$ & $15$ & $15$ & $63$ & $63$\\
    $\kappa$& $0.3528$ & $0.3528$ & $0.4693$ & $0.4744$ & $0.56$ & $0.5608$ & $0.6581$ & $0.6460$\\
    $a$& $1.1327$ & $1.1327$ & $1.1397$ & $1.1234$ & $1.347$ & $1.3452$ & $1.5255$ & $1.5978$\\
    $\alpha$& $1.585$ & $1.585$ & $2.1677$ & $2.0922$ & $5.8884 $ & $5.8264 $ & $18.954 $ & $19.8884$
    \\
    \bottomrule
\end{tabular}
   \label{tab:refined}
\end{table}

\newpage
\begin{table}[htbp]
  \centering
  \caption{Optimized $c_1,c_2$ and $c_3$ parameters for the approximation \eqref{eq:approx_IX} of \cite{BrannstromRasmussenGrant2005}.}

  \begin{tabular}{@{}ccccccccc@{}}
    \toprule
   &  \multicolumn{8}{c}{Modulation Scheme} \\ \cmidrule(l){2-9}
   &  \multicolumn{2}{c}{QPSK}  & \multicolumn{2}{c}{$8$-PSK}  & \multicolumn{2}{c}{$16$-QAM} & \multicolumn{2}{c}{$64$-QAM}\\\cmidrule(l){2-9}
         &   CM   & BICM   &   CM   &  BICM  &   CM   &  BICM  &   CM   &   BICM\\
    $c_1$& $0.77$ & $0.77$ & $0.61$ & $0.81$ & $0.48$ & $0.59$ & $0.47$ & $0.4$\\
    $c_2$& $0.87$ & $0.87$ & $0.68$ & $0.06$ & $0.61$ & $0.06$ & $0.44$ & $0.05$\\
    $c_3$& $1.16$ & $1.16$ & $1.45$ & $1.75$ & $1.48$ & $1.65$ & $1.87$ & $1.63$\\
    \midrule
    $\Delta R$& $0.0033$ & $0.0033$ & $0.0241$ & $0.0223$ & $0.0414$ & $0.0259$ & $0.0977$ & $0.0656$
    \\
    \bottomrule
\end{tabular}
   \label{tab:approx_IX}
\end{table}

\newpage
\begin{table}[htbp]
  \centering
  \caption{Power delay profile of the normalized ETSI BRAN-A channel model using a zero-hold order filter.}

  \begin{tabular}{@{}cc@{}}
    \toprule
    Delay (multiples of $50$ns) & Normalized path power (dB)\\\midrule
    $1$& $-3.4630$\\
    $2$& $-4.6006$\\
    $3$& $-8.9151$\\
    $4$& $-12.8223$\\
    $5$& $-19.9222$\\
    $6$& $-21.1202$\\
    $7$& $-25.4329$\\
    $8$& $-29.7891$\\
    $9$& $-34.1993$
    \\
    \bottomrule
\end{tabular}
   \label{tab:brana}
\end{table}


\newpage

\newpage

\newpage

\newpage

\newpage

\newpage

\newpage

\newpage

\newpage
\begin{figure}
\begin{center}
\includegraphics[width =1\columnwidth]{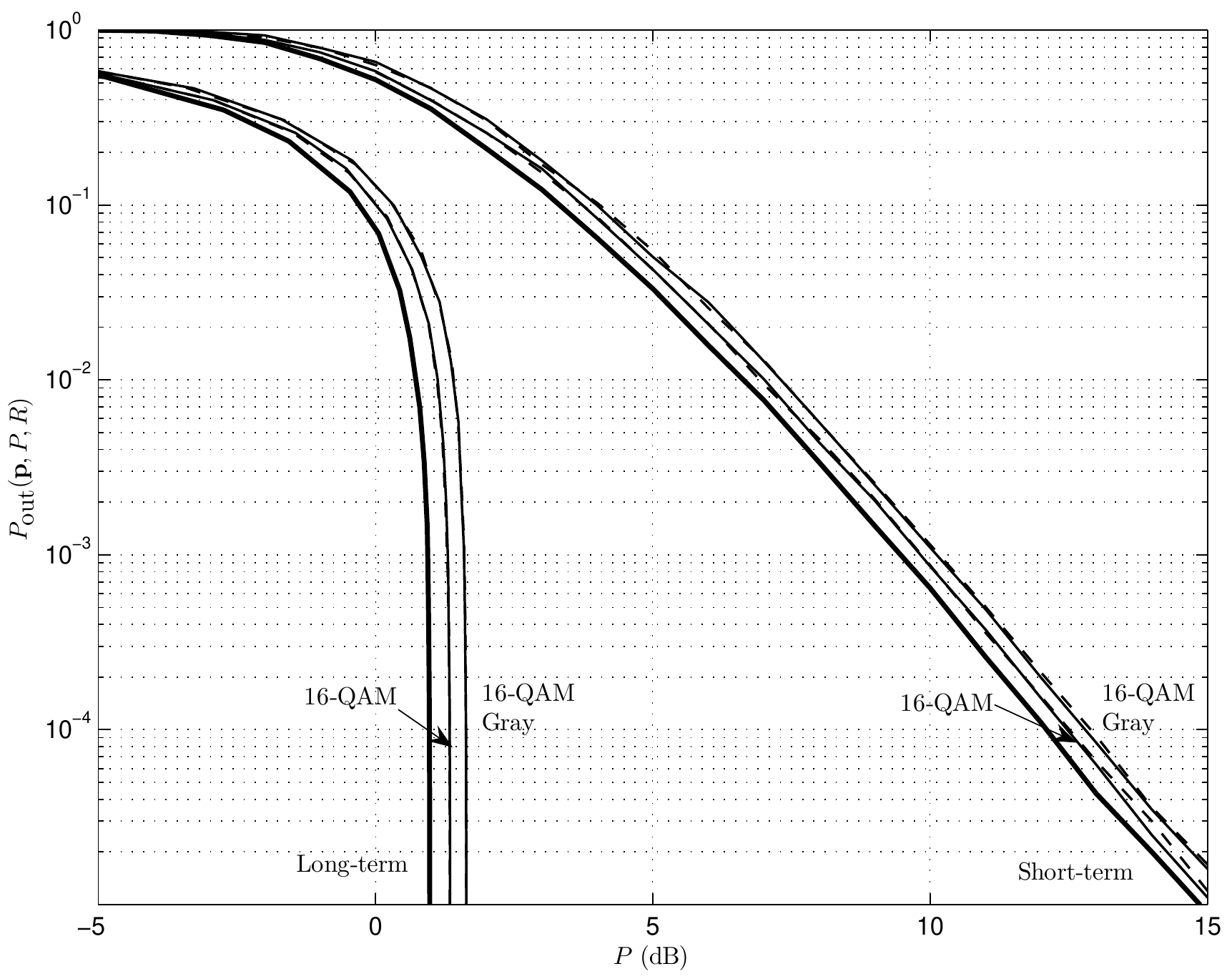}
\caption{Outage performance of various short- and long-term power allocation
schemes in a block-fading channel with $B=4$, $R=1$, Rayleigh fading, Gaussian
and $16$-QAM CM and BICM (with Gray mapping). The thick solid lines correspond
to the Gaussian input; the thin solid-lines
  represent optimal scheme; the dashed lines
  represent truncated water-filling ($\mathbf{p}_{\rm lt}^{\rm tw}(\boldgamma)$
  with $\beta=13$ dB).}
\label{fig: comp_short_long_QAM_BICM}
\end{center}
\end{figure}

\newpage


\end{document}